\documentclass{aa}
\usepackage{hyperref}
\usepackage{txfonts}
\usepackage{graphicx}
\usepackage[ ]{natbib} 
\usepackage{multirow}
\bibpunct{(}{)}{;}{a}{}{,} % to follow the A&A style
\usepackage{color}

\begin{document}

\title{A possible cyclotron resonance scattering feature near 0.7 keV
 in X1822-371}
  
%\subtitle{}

\author{R. Iaria\inst{1}, T. Di Salvo\inst{1}, M. Matranga\inst{1}, 
 C. G. Galiano\inst{1}, A. D'A\'\i\inst{1},  A. Riggio\inst{2}, 
L. Burderi\inst{2}, A. Sanna\inst{2}, C. Ferrigno\inst{3}, M. Del
Santo\inst{4}, F. Pintore\inst{2}, 
   N. R. Robba\inst{1}}

\offprints{R. Iaria, \email{rosario.iaria@unipa.it}}

\institute{Dipartimento di Fisica e Chimica,
Universit\`a di Palermo, via Archirafi 36 - 90123 Palermo, Italy
\and 
Dipartimento di Fisica, Universit\`a degli Studi di Cagliari, SP Monserrato-Sestu, KM 0.7, Monserrato, 09042 Italy
\and 
ISDC, Department of Astronomy,  Universit\'e de Gen\`eve, chemin d'\'Ecogia, 16, CH-1290 Versoix, Switzerland
\and 
Istituto Nazionale di Astrofisica-IAPS, via del Fosso del Cavaliere 100, 00133 Roma} 

\date{}

\abstract
{The source X1822-371 is a low-mass X-ray binary system (LMXB) viewed
  at a high inclination angle.  It  hosts a neutron star with a spin period of
  $\sim 0.59$ s, and recently, the spin period derivative was
  estimated to be $(-2.43 \pm 0.05) \times 10^{-12}$ s/s.}
{Our aim is to address the origin of the large residuals below 0.8 keV
  previously observed in the XMM/EPIC-pn spectrum of X1822-371. }
{We analyse all available X-ray observations of
  X1822-371 made with XMM-Newton, Chandra, Suzaku  and INTEGRAL satellites.
  The observations were not simultaneous.  The Suzaku and INTEGRAL
    broad band energy coverage allows us to constrain the spectral shape of  the
  continuum emission well. We use the model already proposed for
  this source, consisting of a Comptonised component absorbed by
  interstellar matter and partially absorbed by local neutral matter, and we
  added a Gaussian feature in absorption at $\sim 0.7$ keV.  This addition significantly  improves the fit 
   and flattens the residuals between 0.6 and 0.8
  keV. }
{We interpret the Gaussian feature in absorption as a cyclotron
  resonant scattering feature (CRSF) produced close to the neutron star surface and
derive  the magnetic field strength 
  at the surface of the neutron star,  $(8.8 \pm 0.3) \times
  10^{10}$ G for a radius of 10 km. We derive the pulse period in the EPIC-pn data to be 
   0.5928850(6) s and estimate that
  the spin period derivative of X1822-371 is $(-2.55 \pm 0.03) \times
  10^{-12}$ s/s using all available pulse period measurements.  Assuming that the intrinsic luminosity of X1822-371
  is at the Eddington limit and using the values of spin period and
  spin period derivative of the source, we constrain the neutron star
  and companion star masses.  We find the neutron star and the companion
  star masses to be $1.69 \pm 0.13$ M$_{\odot}$ and $0.46 \pm 0.02$
  M$_{\odot}$, respectively, for a neutron star radius of 10 km.}
  % conclusions heading (optional)
{In a self-consistent scenario in which X1822-371 is spinning-up and accretes at the
  Eddington limit, we estimate that the magnetic field of the neutron
  star is $(8.8 \pm 0.3) \times 10^{10}$ G for a neutron star
    radius of 10 km.  If our interpretation is correct, the Gaussian absorption feature near 0.7 keV is 
    the very first detection of a CRSF  below 1 keV in a LMXB. }
%\keywords{X-rays: binaries - stars: neutron - accretion - X-rays: stars: 
%individual: \src  }

\authorrunning{R.\ Iaria et al.}

\titlerunning{Determination of the magnetic field at the surface of the neutron star in X1822-371}

\maketitle

%\clearpage
\section{Introduction}
Neutron stars (NS) are thought to be born with magnetic fields
(B-fields) above $\sim$ 10$^{12}$ G. Direct measurements of the
strength of these fields come from the detection of cyclotron resonant
scattering features (CRSFs).  In accreting NS, X-ray pulsations and
CRSFs are common in systems containing a young high-mass companion
(HMXBs), with typical strengths of the NS B-field between
10$^{12}$--10$^{13}$ G, whereas in binary systems containing a
low-mass companion (LMXBs) pulsations are detected only in a small
fraction of systems, and no CRSF has been detected to
date. The most likely explanation is that in such systems, the NS
B-field is sufficiently decayed in the course of its
evolution to a value that, at accretion rates corresponding to
10$^{35}$--10$^{38}$ erg s$^{-1}$, the corresponding magnetospheric
radius becomes smaller than the NS radius.

In LMXBs when pulsations are detected, the inferred NS B-fields 
are of the order of 10$^{8}$--10$^{9}$ G, which is about three orders
of magnitude less than the typical values for HMXBs. Intermediate
values of the NS B-field, between these two
ranges, are uncommonly observed, most probably for evolutionary
reasons.  However, some notable exceptions exist, such as (i) 
the 11 Hz pulsar IGR J17480-2446
\citep{papitto11}, whose NS B-field was estimated in the range 2
$\times$ 10$^{8}$--2.4 $\times$ 10$^{10}$ G; (ii) the 2.1 Hz X-ray pulsar
GRO J1744-28, with an estimated NS B-field of $\sim$ 2.4 $\times$
10$^{11}$ G \citep{cui97}; and, finally, (iii) the 1.7 Hz X1822-371
\citep{Jonker_2001}.

Despite the small sample that it belongs to, the peculiarity of
X1822-371 still stands out.  
Analysing the RXTE data of
X1822-371, \cite{Jonker_2001} detected for the first time a coherent
pulsation at 0.593 s associated with the NS spin period and
inferred a spin period derivative of $(-2.85 \pm 0.04) \times
10^{-12}$ s s$^{-1}$. Analysing RXTE data from 51~976 to
52~883 MJD, \cite{Jain_2010} constrained the spin period
derivative better, finding $(-2.481 \pm 0.004) \times 10^{-12}$ s
s$^{-1}$. The ephemeris of X1822-371 has  recently been updated
by \cite{iaria_2011}, who estimated an orbital period of 
5.5706124(7) hr and  
an orbital period derivative of
$(1.51 \pm 0.08) \times 10^{-10}$ s s$^{-1}$ when analysing X-ray data
spanning 30 years. A similar sample of data was analysed by
\cite{Burderi_2010} who obtained an orbital period derivative of $(1.50
\pm 0.07) \times 10^{-10}$ s s$^{-1}$. In an independent paper,
\cite{Jain_2010} obtained an orbital period derivative of $(1.3 \pm 0.3)
\times 10^{-10}$ s s$^{-1}$ from X-ray data and, studying the optical and UV data of X1822-371,
\cite{Bayless2010} derived the new optical ephemeris for the source finding an orbital
period derivative of $(2.1 \pm 0.2) \times 10^{-10}$ s s$^{-1}$.

\cite{Burderi_2010} show that the  orbital-period
derivative is three orders of magnitude larger than what is expected
from conservative mass transfer driven by magnetic braking and/or
gravitational radiation. They conclude that the mass transfer rate
from the companion star is between 3.5 and 7.5 times the Eddington
limit ($\sim 1.1 \times 10^{18}$ g s$^{-1}$ for a NS mass of 1.4
M$_{\odot}$ and NS radius of 10 km), suggesting that the mass transfer has
to be highly non-conservative, with the NS accreting at the
Eddington limit and the rest of the transferred mass expelled from the
system  by the  radiation pressure.  \cite{Bayless2010} show that the accretion rate onto the
NS should be $\sim 6.4 \times 10^{-8}$ M$_\odot$ yr$^{-1}$ in a
conservative mass transfer  scenario, again suggesting a highly non-conservative
mass transfer.

The large orbital period derivative is a clear clue that the intrinsic
luminosity of X1822-371 is at the Eddington limit, which is almost two orders of
magnitude higher than the observed luminosity \citep[i.e. $\sim
10^{36}$ erg s$^{-1}$, see
e.g.][]{Hellier_mason1989,Heinz_nowak_2001,parmar2000,Iaria2001_1822}.
This is also supported by the ratio $L_X/L_{opt}$ of X1822-371.
\cite{Hellier_mason1989} showed that the ratio $L_X/L_{opt}$ for
X1822-371 is $\sim 20$, a factor 50 smaller than the typical value of
1000 for the other LMXBs.  This suggests that the intrinsic X-ray
luminosity is underestimated by at least a factor of 50.
Finally, \cite{Jonker_2001} show that for a luminosity  of $10^{36}$ erg s$^{-1}$, 
the NS B-field  strength 
assumes an unlikely value of $8 \times 10^{16}$ G,
while for a luminosity of the source of $\sim 10^{38}$  erg s$^{-1}$,
 it assumes a more conceivable value of $8 \times 10^{10}$ G.

 Recently when analysing an XMM-Newton observation of X1822-371
 and using RGS and EPIC-pn data, \cite{iaria_2013} fitted the X-ray spectrum of this source to a model
 consisting of a Comptonised component {\tt CompTT}\footnote{The
   names of the cited spectral models are consistent with
   those adopted in the spectral fitting package XSPEC
   \citep{Arnaud_96}. The models are described in
   \url{http://heasarc.gsfc.nasa.gov/xanadu/xspec/manual/XspecModels.html}}
 absorbed by interstellar neutral matter and partially absorbed by
 local neutral matter. The authors took the Thomson scattering of the
 local neutral matter into account by adding the {\tt
   cabs}\footnotemark[1] component and imposed that the equivalent
 hydrogen column density of the {\tt cabs} component is the same as
 the local neutral matter. The adopted model is similar to the one
 previously used by \cite{Iaria2001_1822} to fit the averaged BeppoSAX
 spectrum of X1822-371.   \cite{iaria_2013} suggest that the
 Comptonised component is produced in the inner regions of the system,
 which are not directly observable. The observed flux is only 1\%
 of the 
 total intrinsic luminosity, the fraction
  scattered along the line of sight by an
 extended optically thin corona with an optical depth $\tau \simeq
 0.01$. This scenario explains why the observed luminosity of the
 source is $\sim 10^{36}$ erg s$^{-1}$, while the 
 orbital  period derivative suggests 
  an intrinsic luminosity of X1822-371 at the Eddington limit.
 Furthermore, \cite{iaria_2013} found that large residuals are present
 in the EPIC-pn spectrum below 0.9 keV and fitted those residuals by adding a
 black-body component with a temperature of 0.06 keV, although they
 suggest that further investigations were needed to understand the
 physical origin of this component.

 Recently, \cite{sasano_13} have analysed a Suzaku observation of X1822-371
 in the 1-45 keV energy range. The authors detect the NS pulsation
 in the HXD/PIN instrument at 0.5924337(1) s and inferred a spin
 period derivative of $(-2.43 \pm 0.05) \times 10^{-12}$ s/s.
 \cite{sasano_13} also suggest the presence of a CRSF at 33 keV and
 inferred from this value a NS B-field of $\sim 2.8 \times 10^{12}$ G
 and a luminosity of the source of $\sim 3 \times 10^{37}$ erg
 s$^{-1}$.

 In this work we determine the NS spin period of X1822-371
   during the XMM-Newton observation and derive a new estimation of the spin period
   derivative,  also taking all the measurements of the spin
   period reported in literature into account, including our derived value. We analyse the combined spectra of
   X1822-371 obtained with XMM-Newton, Chandra \citep[the same data sets
as   analysed by][]{iaria_2013}, Suzaku \citep[the same data set as
   analysed by][]{sasano_13}, and INTEGRAL. We show the presence of
   large residuals close to 0.7 keV,  while we do not find evidence of  a
   cyclotron feature at 33 keV, unlike what has been suggested
   by \cite{sasano_13}. Moreover, we show that a CRSF at 33 keV would
   not be consistent with the evidence that the NS in X1822-371 is
   spinning up.  Fitting the residuals near 0.7 keV with a CRSF
   centered at 0.72 keV, we determine a NS B-field (surface) strength between  $
   7.8 \times 10^{10}$ and $ 9.3 \times 10^{10}$G for a NS radius ranging between 9.5 and
   11.5 km.

 \section{Observations}

\subsection{The Suzaku observation}

The X-ray satellite Suzaku observed X1822-371 on 2006 October 2 with
an elapsed time of 88 ks.  Both the X-ray Imaging Spectrometers
(0.2-12 keV, XISs; \citealt{koyama_07}) and the Hard X-ray Detector
(10-600 keV, HXD; \citealt{taka_07}) instruments were used during
these observations. There are four XIS detectors, numbered as 0 to 3.
XIS0, XIS2, and XIS3 all use front-illuminated CCDs and have very
similar responses, while XIS1 uses a back-illuminated CCD.  The HXD
instrument includes both positive intrinsic negative (PIN) diodes
working between 10 and 70 keV and the gadolinium silicate (GSO)
scintillators working between 30-600 keV. Both the PIN and GSO are
collimated (non-imaging) instruments.  During the observation, XIS0 and
XIS1 worked in 1/4 Window option, while XIS2 and XIS3 worked in full
window.  The effective exposure time of each XIS CCD is nearly 38 ks,
and the HXD/PIN exposure time is nearly 31 ks. 

We reprocessed the data using the {\tt aepipeline} tool
provided by Suzaku FTOOLS version 20\footnote{See
  \url{http://heasarc.gsfc.nasa.gov/docs/suzaku/analysis/suzaku_ftools.html}
  for more details} and applying the latest calibration available as
of 2013 November.  We then applied the publicly available tool {\tt
  aeattcor.sl}\footnote{
  \url{http://space.mit.edu/ASC/software/suzaku/aeatt.html} } by John
E. Davis to obtain a new attitude file for each observation. This tool
corrects the effects of thermal flexing of the Suzaku spacecraft and
obtains a more accurate estimate of the spacecraft attitude.  For  
our observation, the above attitude correction produces sharper
point-spread-function (PSF) images. With the new attitude file, we
updated the XIS event files using the FTOOLS {\tt xiscoord} program.
We estimated the pile-up fractions using the publicly available tool
{\tt pileup\_estimate.sl}\footnote{
  \url{http://space.mit.edu/ASC/software/suzaku/pile_estimate.sl} } by
Michael A. Nowak. The pile-up fraction refers to the ratio of events
lost via grade or energy migration to the events expected in the
absence of pile-up. The unfiltered pile-up fractions integrated over a
circular region centred on the brightest pixel of the CCD and with a
radius of 105$\arcsec$  are 4.6\%, 3.9\%, 10.4\%, and 9.9\% for XIS0,
XIS1, XIS2, and XIS3, respectively. The large pile-up fraction in XIS2
and XIS3 is due to the two CCDs working in full window during the
observation. To mitigate the pile-up effects in the spectra extracted
from XIS2 and XIS3, we used annular regions, while we adopted circular
regions with a radius of 105$\arcsec$ to extract the spectra from XIS0
and XIS1.  Adopting  annulus regions with inner and outer radii of
28$\arcsec$ and 105$\arcsec$, respectively,  the pile-up
fractions are 4.8\% for XIS2 and 4.6\% for XIS3.  The
background spectra were extracted using the same regions as were adopted to
extract the source spectra and centred where the influence of the
source photons is weak (or absent) in the CCDs.  The response files
of the XIS for each observation were generated using the {\tt
  xisrmfgen} Suzaku tool, and the corresponding ancillary files were
extracted using the {\tt xisarfgen} Suzaku tool, suitable for 
 a point-like source.  Because the responses of XIS0, XIS2, and XIS3 are on
the whole very similar, we combined their spectra and responses
using the script {\tt addascaspec}.  The XIS spectra were rebinned to have 1024 energy channels.

We also extracted the PIN spectra using the Suzaku tool {\tt
  hxdpinxbpi}.  The non X-ray and cosmic X-ray backgrounds were taken
into account. The non X-ray background (NXB) was calculated from the
background event files distributed by the HXD team. The cosmic X-ray
background (CXB) is from the model by \cite{Bold_87}. The response
files provided by the HXD team were used. The GSO data were not used,
considering the low signal-to-noise ratio above 40 keV.

\subsection{The XMM-Newton observation}

The region of the sky containing X1822-371 was observed by XMM-Newton
between 2001 March 07 13:12:48 UT and March 08 03:32:53 UT
(Obs. ID. 0111230101) for a duration of 53.8 ks.  The European Photon
Imaging Camera (EPIC) on-board XMM-Newton consists of three co-aligned
high-throughput X-ray telescopes.  Imaging charge-coupled-device (CCD)
detectors were placed in the focus of each telescope.  Two of the CCD
detectors are Metal Oxide Semi-conductor (MOS) CCD arrays
\citep[see][]{turner_01}, while the third camera uses pn CCDs
\citep[hereafter EPIC-pn, see][]{struder_01}.  Behind the two
telescopes that have the MOS cameras in the focus, about half of the
X-ray light is utilised by the reflection grating spectrometers
(RGSs). Each RGS consists of an array of reflection gratings that
diffracts the X-rays to an array of dedicated CCD detectors
\citep[see][]{brinkman_98,denherder_01}.

 During the
observation, MOS1 and MOS2 camera were operated in fast uncompressed
mode and small window mode, respectively.  The EPIC-pn camera was operated
in timing mode with a medium filter during the observation.  The faster
CCD readout results in a much higher count rate capability of 800
cts/s before charge pile-up become a serious problem for point-like
sources. The EPIC-pn count rate of the source was around 55 cts/s, thereby
avoiding telemetry and pile-up problems.

Although the  RGS and EPIC-pn data products 
were extracted and analysed by \cite{iaria_2013}, we
 extracted the data products of the RGS and EPIC-pn camera again using the
very recent science analysis software (SAS) version 13.5.0 and the
calibration files available on 2013 Dec. 17.  
We used the SAS tools {\tt rgsproc}, {\bf {\tt emproc,}} and {\tt epproc} to obtain the RGS, MOS, and EPIC-pn data products.

Since the EPIC-pn was operated in timing mode during the observation,
we extracted the EPIC-pn image of RAWX vs. PI to select appropriately the
source and background region. The source
spectrum is selected from a box region 
 centred on RAWX=38 with a  width of 18
columns. The background spectrum was selected from a box region
 centred on RAWX=5 with a width of two
columns.  We extracted only single and double events (patterns 0 to 4)
for the source and background spectra and applied the SAS tool {\tt
  backscale} to calculate the different areas of the source and background
regions.

We extracted the MOS1 source spectrum, adopting a box region centred
on RAWX=317 having a width of 50 pixels.  The MOS1
background spectrum is extracted from a source-free region selected
in one of the outer CCDs that collect photons in imaging mode during the
observation.  We extracted the MOS1 spectrum and the corresponding
redistribution matrix and ancillary files using the standard
recipe\footnote{ see
  \url{http://xmm.esac.esa.int/sas/current/documentation/threads/MOS_spectrum_timing_thread.shtml}}.
We also extracted the source+background light curve, observing that the
average count rate during the observation is 15 c/s.  The MOS2
source spectrum was extracted from a circular region centred on the
pixel showing the largest number of photons; the radius of the region
is 640 pixels. A circle with radius of 640 pixels was placed in a
source-free region to extract the background spectrum.  We extracted
the MOS2 spectrum and the corresponding redistribution matrix and
ancillary files using the standard recipe.\footnote{ see
  \url{http://xmm.esac.esa.int/sas/current/documentation/threads/MOS_spectrum_thread.shtml}}
We also extracted the background-subtracted light curve using the SAS
tool {\tt Epiclccorr}, and the average count rate is 15 c/s.  Since the
 count-rate limit for avoiding pile-up for the MOS cameras is 100 c/s and 5
c/s for a point-like source in timing uncompressed and small
window modes\footnote{ see
  \url{http://xmm.esac.esa.int/external/xmm_user_support/documentation/uhb_2.1/node28.html}},
respectively, we expect that the pile-up effects are present in the
MOS2 spectrum.
 The RGS1, RGS2, MOS1, MOS2, and EPIC-pn spectra have an
exposure time of 53, 51, 51, 51, and 51 ks, respectively.

\subsection{The Chandra observations}
The Chandra satellite observed X1822-371 from 2008 May 20
22:53:00 to 2008 May 21 17:06:51 UT (Obs. ID. 9076) and from 2008 May
23 13:20:56 to 2008 May 24 12:41:54 UT (Obs. ID. 9858) using the HETGS
for total observation times of 66 and 84 ks, respectively. Both
observations were performed in timed faint mode.  The two Chandra
observations had already been analysed by \cite{iaria_2013}.  In this work
we only used the first-order MEG spectrum \citep[see][for details on the data
extraction]{iaria_2013} because the first-order HEG spectrum has a much lower
effective area  with respect to MEG below 1 keV, and its analysis is not useful for
investigating the presence of features in the spectrum of X1822-371 at
those energies. The first-order MEG spectrum has a total exposure time of
142 ks.

\subsection{The INTEGRAL observations}
The INTErnational Gamma-Ray Astrophysics Laboratory
\citep[INTEGRAL][]{winkler_2003} has repeatedly observed the
X1822-371 region.  We searched the whole IBIS
\citep{ubertini_2003} and JEM-X \citep{lund_2003} public  catalogues, selecting 
 only pointings (science windows, SCW) with sources
  within six degrees of the centre of the field of view and with
  exposures longer than 500 s to reduce the calibration uncertainties
  of the IBIS/ISGRI \citep{lebrun_2003} spectral response.
The available IBIS
  data set covers the period starting from 2003 March 21 until 2013 March 21
  for a total usable on-source time of 1~271 ks and an effective
  dead-time corrected exposure of 874 ks.  Because of the smaller
  field of view, the total exposure of JEM-X1 (camera 1) is 330 ks,
  while the dead-time-corrected exposure is 283 ks.  The INTEGRAL data
  analysis uses standard procedures within the offline
  science analysis software (OSA10.0) distributed by the ISDC
   \citep{courvoisier_2003}.  In the catalogue used for
  the extraction of the IBIS spectra, we  have included all the sources significantly
  detected  in the total image obtained by
  mosaicking the individual pointings.  We exploited a custom
  spectral binning optimised in the energy range 20-100 keV, and in
  the detection of  spectral features around 30 keV, weighted the
  time-evolving response function according to the available data and
  excluded the data below 21 keV due to the evolving detector's
  low threshold.  For JEM-X, we adopted the standard 16 bins  spectrum
  provided by the analysis software and excluded the data below 5 keV
  and above 22 keV, which are affected by calibration uncertainties.
  \footnote{The JEM-X2 unit was active only during a limited part of
    the mission, so the exposure time is not enough to provide a
    significant spectral constraint.}

\section{Search for the spin period in the XMM data}
\label{timing}
\begin{figure}[ht]
\resizebox{\hsize}{!}{\includegraphics{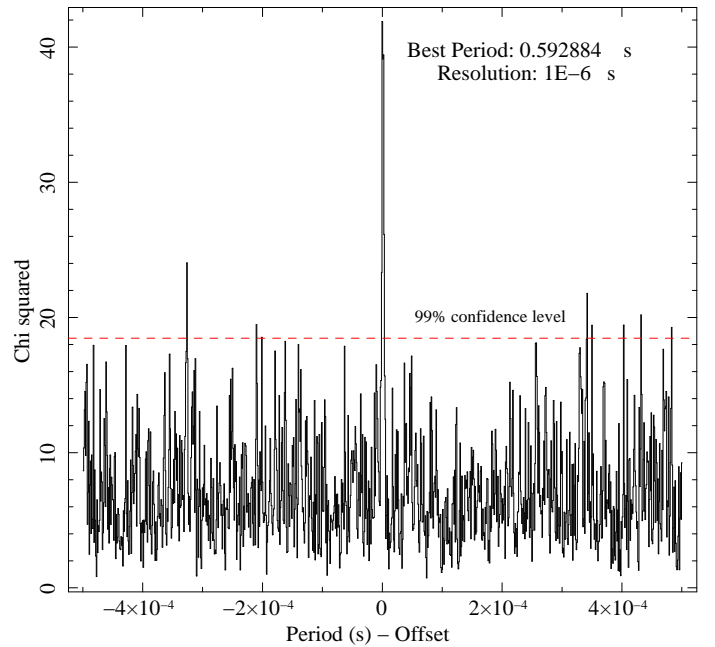}}
\caption{Folding search for periodicities in the 2-5.4 keV EPIC-pn
  light curve. We adopt 8 phase bins per period for the trial-folded
  light curves and a resolution of period search of $1 \times 10^{-6}$
  s. The peak of $\chi^2$ is detected at 0.592884 s. The horizontal
  dashed line indicate the $\chi^2$ value of 18.47 at which we have the
  99\% confidence level for a single trial. }
\label{efsearch}
\end{figure}

We used the EPIC-pn events to search for the spin period.  We applied
the barycentric correction with respect to the source coordinates,
 given by \cite{iaria_2011}, using the SAS tool {\tt barycen};
subsequently, we corrected the data for the orbital motion of the binary
system using the recent X-ray ephemeris of X1822-371 derived by
\cite{iaria_2011} (see Eq. 2 in that work) and the $a  \sin i $ value 
of 1.006 lt-s \citep[see Table 1 in ][]{Jonker_2001}. We selected the EPIC-pn
events in the 2-5.4 keV energy range and explored the period window
between 0.592384 and 0.593384 s using the FTOOL {\tt efsearch} in the
XRONOS package. We adopted eight phase bins per period (a bin time
close to 0.074 s) for the trial folded light curves and a resolution
of the period search of $1 \times 10^{-6}$ s. We observed a $\chi^2$ peak of
41.89  at 0.592884 s, as shown in Fig. \ref{efsearch}. We fitted
the peak with a Gaussian function, assumed the centroid of the
Gaussian as the best estimation of the spin period, and associated the error 
 derived from the best fit.   We find that the
spin period during the XMM observation is 0.5928850(6) s,  and the associated
error is at the 68\% confidence level.

Considering that we have seven degrees of freedom, the probability of
obtaining a $\chi^2$ value greater than or equal to 41.89 by chance is $5.47
\times 10^{-7}$ for a single trial.  In our search we adopted $10^{3}$
trials (we span $10^{-3}$ s with a resolution of the period search of $1
\times 10^{-6}$), consequently we expect  a number of $\simeq 5.5
\times 10^{-4}$ periods with a $\chi^2$-value greater than or equal to
41.89. This implies that our detection is significant at the 99.945\%
confidence level.

Then, we folded the 2-5.4 keV EPIC-pn light curve using the spin period
of  0.5928850(6) s and adopting 16 phase bins per period. We used the arbitrary value of  51975.85 MJD
as epoch of reference. 
\begin{figure}[ht]
\resizebox{\hsize}{!}{\includegraphics{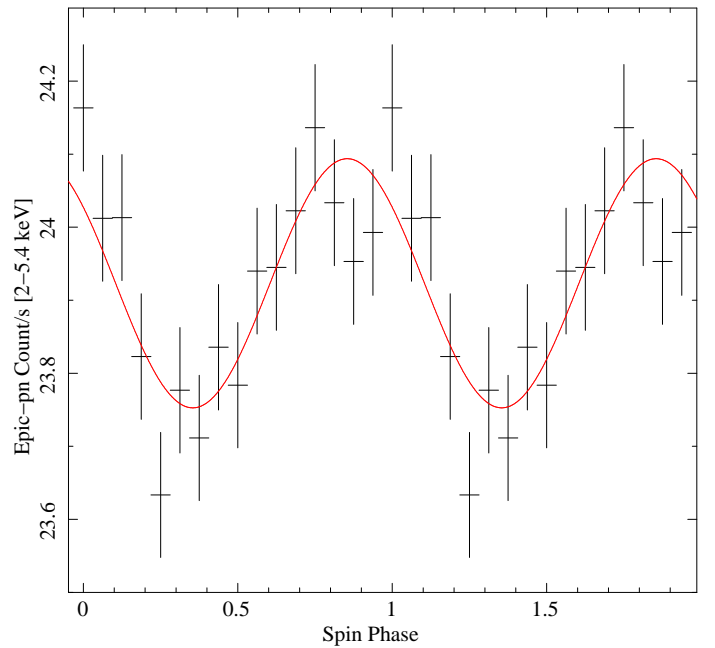}}
\caption{EPIC-pn folded light curves of X1822-371 using the folding period 
 0.5928850(6) s. The folded light curve is obtained using 16 phase bins
per period. }
\label{folded}
\end{figure}
The folded light curve is almost sinusoidal (see Fig.  \ref{folded}).
Fitting the folded light curve with a constant plus a sinusoidal
function with period fixed at one, we obtain a $\chi^2$(d.o.f.) of
13.9(13). Since the constant is 23.92(2) c/s, the background count
rate is close to 1.3 c/s and the amplitude of the sinusoidal function
0.17(3) c/s, we estimate that the pulse fraction is $0.75 \pm 0.13$\%
compatible within 3$\sigma$ to the value of $0.25 \pm 0.06$\% reported
by \cite{Jonker_2001} using RXTE/PCA data in the same energy band.

We report in Table \ref{mideclipse} the 13 values of the spin period
of X1822-371 and the corresponding errors previously estimated,
together with the value found in the present work.  The corresponding
 times are the mean values between the start and stop time 
of the observations in which the spin period was detected, 
the associated errors are one half of the  duration of the corresponding
observation. After deriving the geometric mean of the times in 
  Table \ref{mideclipse} (column 1) obtaining $T_{mid}= 52,257.78 $ MJD,
\begin{table}[ht]
  \caption{Times and corresponding
spin periods}
% title of Table
\label{mideclipse}      % is used to refer this table in the text
%\centering                                      % used for centering table
\begin{center}
\begin{tabular}{ c c c}          % centered columns (4 columns)
\hline\hline                        % inserts double horizontal lines

Times& Spin period & Ref. \\
  (MJD) & (s) & \\
\hline

  50,352.9(6)  &   0.59325(2) & 1\\
50,993.4(6)  &   0.59308615(5) & 1 \\

51,975.9(3)  &    0.5928850(6) & 2 \\

51,976.04(6)   &    0.59290132(11) & 3 \\      
  52,094.80(6)    &    0.59286109(8)  &3 \\
   52,095.73(6)     &     0.59286421(12)& 3\\
 52,432.62(18)      &      0.5927922(13)&3 \\       
  52,489.7(9)     &         0.5927790(6)&3 \\ 
  52,491.61(15)      &      0.5927795(11) &3 \\
  52,503.45(12)      &      0.5927737(10) &3 \\ 
  52,519.41(18)       &      0.5927721(8)  &3 \\ 
  52,882.15(9)       &      0.5926793(15) &3 \\    
 52,883.15(9)       &      0.5926852(21) &3 \\  
 54,010.0(6)          &      0.5924337(10)&4\\

\hline                                             %inserts single line
\end{tabular}
\end{center}

{\small \sc Note} \footnotesize---References: 1 \cite{Jonker_2001},
2 this work, 3 \cite{Jain_2010}, and 4 \cite{sasano_13}.  
\end{table} 
we fitted the spin periods with respect to the  times with $T_{mid}$ subtracted using a
linear function to estimate the spin period derivative.  Unlike
\cite{Jain_2010}, we did not obtain a good fit, 
since the $\chi^2$(d.o.f.)
was close to $10^5$(12); however, we obtained  a very high  value of
-0.9992 of the Pearson  correlation coefficient.
The reason for the high $\chi^2$ value is not clear to us, 
but it could be due to an underestimation of the errors (e.g. small differences
in the orbital ephemeris used to correct the data) or model complications 
(e.g.  caused by small fluctuations around an average linear trend),
 or other issues. The detailed investigation of this aspect goes beyond 
the aim of this paper.

\begin{figure}[ht]
\resizebox{\hsize}{!}{\includegraphics{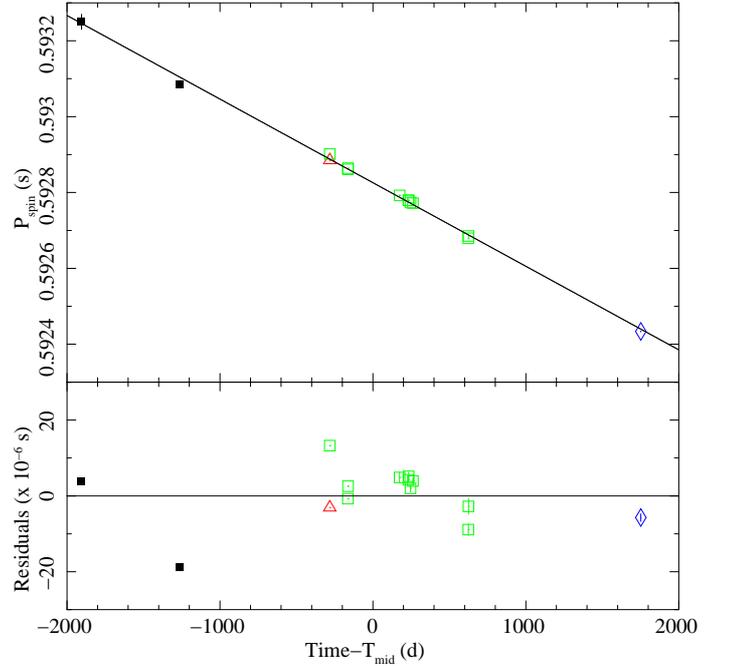}}
\caption{Top panel: spin period values shown in
  Table \ref{mideclipse} vs.  time in units of days (see the text). The linear best fit is also plotted.
  The black squares, open green squares, blue diamond, and red
  triangle indicate the spin period values reported by
  \cite{Jonker_2001}, \cite{Jain_2010}, \cite{sasano_13}, and this
  work, respectively.  Bottom panel: the corresponding residuals in
  units of $10^{-6}$ s.}
\label{data}
\end{figure}

Considering the high value of the $\chi^2$ for the linear fit and to
 estimate the  error associated  to the spin period derivative, 
 we fitted the 14 points without the estimated errors
and attributed the post-fit errors to the best-fit parameters under the
assumption that the model is reliable. 
In this way we at least get an estimation of the averaged linear trend of the measured spin 
period with respect to time.
Fitting the data again with a linear  function, we obtain $a = 0.592826(6)$ s and $b =
-2.20(3) \times 10^{-7}$ s/d, with the errors at 68\% confidence
level. This implies that the spin period derivative, $\dot{P_s}$, is
$-2.55(3) \times 10^{-12}$ s/s;
this is compatible within three
sigmas with the previously reported values of $-2.481(4) \times
10^{-12}$ s/s and $-2.43(5) \times 10^{-12}$ s/s  given by \cite{Jain_2010}
and \cite{sasano_13}, respectively. We show in Fig. \ref{data} (top
panel) the 14 points and the corresponding linear best fit.
The
corresponding residuals are shown in Fig. \ref{data} (bottom panel).

Furthermore, we  search for the same periodicity in the MOS1 data
(taken in timing mode); unfortunately, the lower statistics  with respect to the 
EPIC-pn data do not allow us to detect the periodicity in this
data set.

\begin{figure}[ht]
\resizebox{\hsize}{!}{\includegraphics{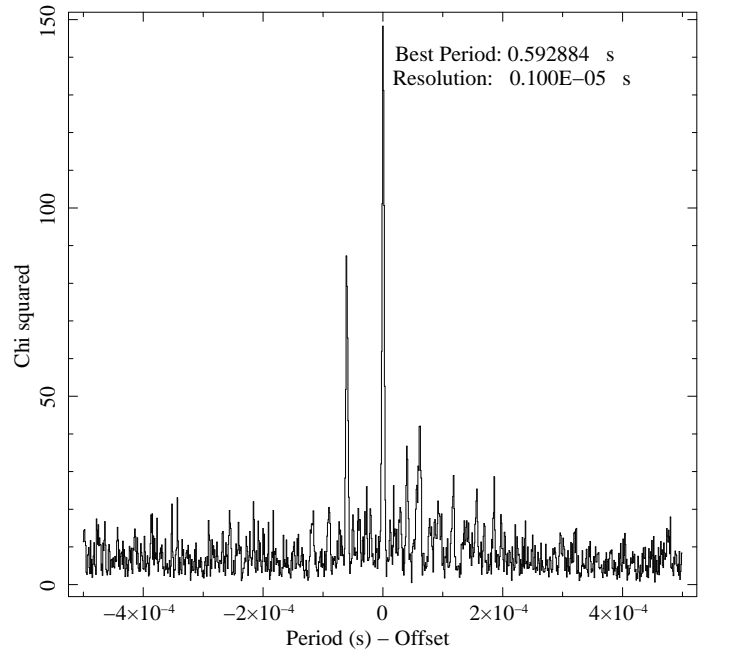}}
\caption{Folding search  for periodicities in the RXTE/PCA observations.  
The peak of $\chi^2$ is detected at 0.592884 s. }
\label{efsearch_xte}
\end{figure}

We note that the spin period obtained from the EPIC-pn
  events is inconsistent with the value reported by \cite{Jain_2010}
  using RXTE/PCA observations simultaneous to the XMM observation that we
  analyse in this work (see Table \ref{mideclipse} and
  Fig. \ref{data}). To confirm the robustness of our results, we
  reanalysed the simultaneuous  RXTE/PCA observations (P50048-01-01-00,
  50048-01-01-01, P50048-01-01-02, P50048-01-01-03, P50048-01-01-04,
  P50048-01-01-05, P50048-01-01-06) spanning 2001 March 7 10:29:26
  UT to 2001 March 8 4:47:44 UT.
We applied
the barycentric correction with respect to the source coordinates
and  corrected the data for the orbital motion of the binary
system as done for the EPIC-pn data.   No energy selection was applied
to the RXTE/PCA events file.  
 We explored the period window
between 0.592384 and 0.593384 s using the FTOOL {\tt efsearch} in the
XRONOS package. We adopted eight phase bins per period for the trial-folded light curves and a resolution
of the period search of $1 \times 10^{-6}$ s. We observe a $\chi^2$ peak of
150  at 0.592884 s, as shown in Fig. \ref{efsearch_xte}. We fitted
the peak with a Gaussian function, assumed the centroid of the
Gaussian as the best estimation of the spin period, and associated the error 
 derived from the best fit.   We find that the
spin period during the RXTE/PCA observations is 0.5928846(3) s and that the associated
error is at the 68\% confidence level.
This result confirms our detection in the XMM/Epic-pn data.
Finally, we note that the value of the spin period derivative obtained above
does not significantly change when using our value instead of the one reported in
Table \ref{mideclipse}.

\section{Energy range selection for the  XMM-Newton, Suzaku, Chandra, and INTEGRAL spectra} 
\label{selection} 
 We rebinned the MOS1 and MOS2  spectra using the SAS tool {\tt specgroup}
 to have at least 25 counts per energy channel and with an
  over-sample factor of 5.  To verify that the spectra are similar and see how
  the pile-up effects influence the MOS2 spectrum, we fitted 
  the two MOS spectra in the 0.6-10 keV energy range using XSPEC
  (version 12.8.1).  We adopted a very simple model consisting of {\tt
    phabs*pcfabs*CompTT}\footnotemark[1], similar to the model used by
  \cite{Iaria2001_1822} to fit the BeppoSAX broad-band spectrum of
  X1822-371.  The {\tt phabs} component takes the
  photoelectric absorption by the interstellar neutral matter into account, it is a
  multiplicative component defined as $M(E) = \exp[-N_H*\sigma(E)]$,
  where $\sigma(E)$ is the photo-electric cross-section (not including
  Thomson scattering).  The free parameter, $N_H$, is the equivalent
  hydrogen column density in units of 10$^{22}$ atoms cm$^{-2}$.  The {\tt
    pcfabs} component takes the photoelectric absorption into account
  owing to the neutral matter near the source, and it is a multiplicative
  component defined as $M(E) = f*\exp[-N_{H_{pc}}*\sigma(E)] + (1-f)$, where
 $N_{H_{pc}}$ is the equivalent hydrogen column in units of
  10$^{22}$ atoms cm$^{-2}$ of the local neutral matter, and $f$ is a
  dimensionless free parameter ranging between 0 and 1 that takes the fraction of emitting region occulted by the local neutral matter into
  account.
  
\begin{figure}[ht]
\resizebox{\hsize}{!}{\includegraphics{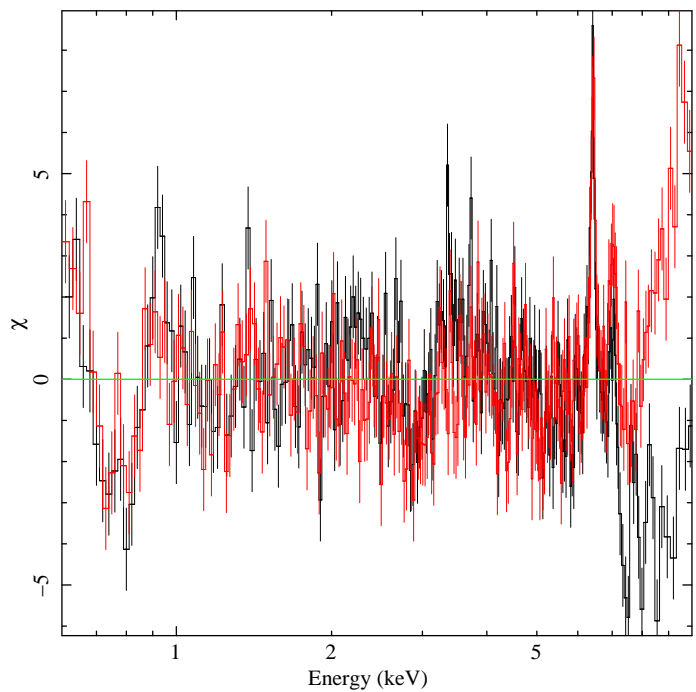}}
\caption{MOS1 (black) and MOS2 (red) residuals with respect to the model described in the text. }
\label{mos1_e_mos2}
\end{figure}

The {\tt CompTT} component \citep{titarciuk_94} is a Comptonisation
  model of soft photons in a hot plasma.  For this component, the soft
  photon input spectrum is a Wien law with a seed-photon temperature,
  $kT_0$, that is a free parameter. The other free parameters of the
  component are the plasma temperature, $kT_e$, the plasma optical
  depth $\tau$, and the normalisation $N_{CompTT}$.  We  used 
   a slab geometry for the Comptonising cloud.  We fitted the
  0.6-10 keV MOS1 and MOS2 spectra simultaneously with the aim of
  estimating the pile-up effects in the MOS2 spectrum.  We obtained a
  large $\chi^2({\rm {d.o.f.}})$ of 1716(540).  The two spectra are consistent 
  with each other
 between 0.6 and 7 keV. Above 7 keV the pile-up distortion
  is evident in the MOS2 spectrum.  Furthermore, the presence of two emission lines in the
  residuals at 6.4 and 6.97 keV is evident;
  finally, both the spectra are not well fitted between 0.6 and 1 keV and 
  show large residuals.  We
  show the MOS1 and MOS2 residuals in
  Fig. \ref{mos1_e_mos2}.
We  sum the MOS1 and MOS2 spectra using a new
recipe\footnote{ see
  \url{http://xmm.esac.esa.int/sas/current/documentation/threads/Epic_merging.shtml}}.
In the following the summed spectrum is called MOS12 spectrum.  
The MOS12 spectrum
is rebinned with an over-sample factor of 5. 
 
We added the first-order
spectrum of RGS1 and RGS2 together using the SAS tool {\tt rgscombine}; 
 hereafter, the summed spectrum is called RGS12. We rebinned the 
 RGS12 to have at least  200 counts per energy channel. In the following, we analyse the RGS12 spectrum in  the 0.35-2 keV energy range.

 %Fine GRASSETTO

\begin{figure}[ht]
\resizebox{\hsize}{!}{\includegraphics{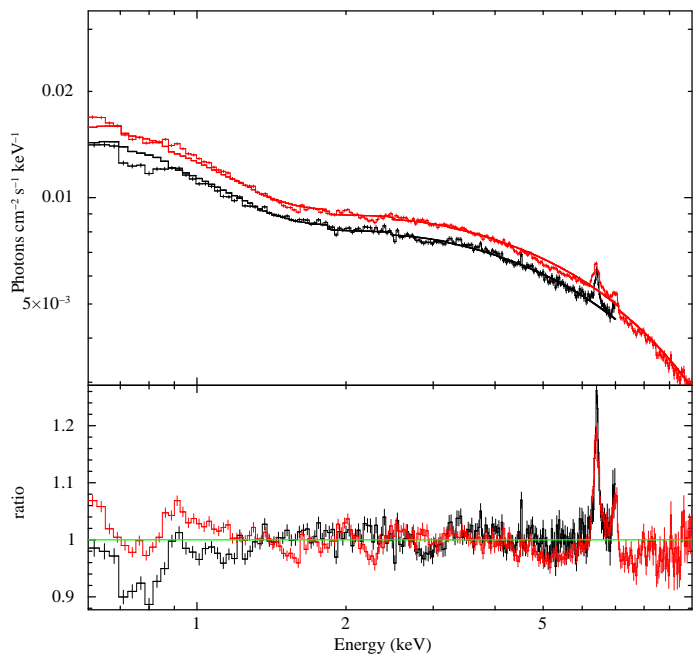}}
\caption{MOS12 (0.6-7 keV; black) and EPIC-pn (0.6-10 keV; red) spectra (top panel) and ratio (data/model; bottom panel) with respect to the model described in the text. }
\label{mos12_Epic}
\end{figure}

The EPIC-pn spectrum is rebinned using the SAS tool
{\tt specgroup} imposing
at least 25 counts per energy channel and an over-sample factor of 5.
%%%%%%%%%%%%%%%%%%%%%%%%%%%%%%%%%%%%%%%%%%%%%%%%%%%%%%%%%%5
% CONFRONTO EPIC MOS12 
 To check the consistency of the EPIC-pn and MOS12 spectra, we fit
  them simultaneously adopting the same model described above.   We
  obtain a large $\chi^2({\rm {d.o.f.}})$ value of $ 3228(466)$.  We
  show the two spectra (MOS12 and EPIC-pn spectra) and the corresponding ratio (data/model) in
  Fig. \ref{mos12_Epic}.
\begin{figure}[ht]
\resizebox{\hsize}{!}{\includegraphics{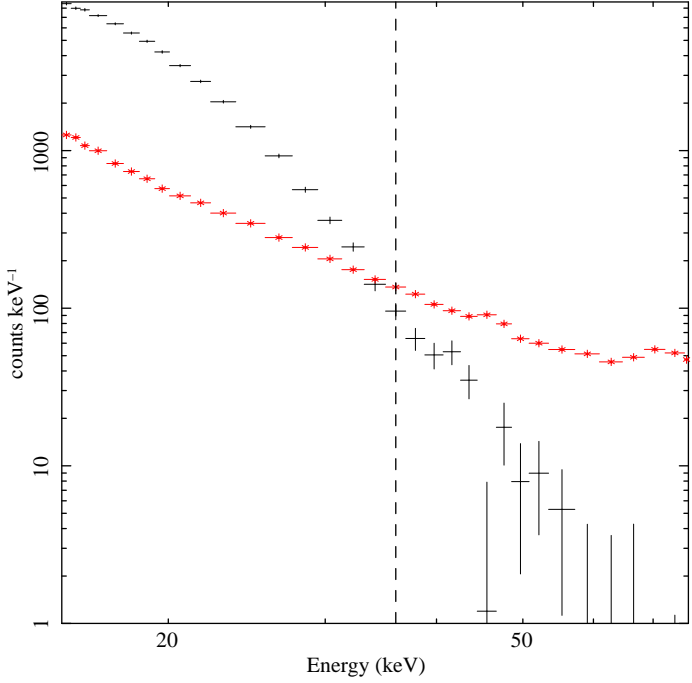}}
\caption{HXD/PIN source spectrum (black) and NXB+CXB 
spectrum  (red) are shown. The NXB+CXB spectrum overwhelms the source 
spectrum at energies larger than 36 keV, the energy threshold is indicated
with a dashed vertical line.   }
\label{backpin}
\end{figure}
We observe a large absorption feature in both spectra at 0.7 keV.  A
mismatch between the two spectra is evident below 1.5 keV; the EPIC-pn
residuals show an instrumental feature at 2.2 keV owing to the neutral
gold M-edge. Finally, we detect the presence of  strong emission lines in the Fe-K
region.  The causes of the mismatch between the
  EPIC-pn and MOS12 are not clear. However,
we note that the residuals close to 0.7 keV are 
  present in both the EPIC-pn and MOS12 spectra (see
  Fig. \ref{mos12_Epic}, lower panel). We adopt the 0.6-10 keV
  energy range for the EPIC-pn spectrum.  To fit the instrumental
  feature at 2.2 keV in the EPIC-pn spectrum, we  use an absorption
  Gaussian line with the centroid and the width fixed at 2.3 and 0
  keV, respectively. We  fit  MOS12 spectrum using 
    the 0.6-7 keV or the 1.5-7 keV energy band.

%To avoid the mismatch between the EPIC-pn and MOS12
%spectrum we will analyse the MOS12 spectrum between 1.5 and 7
%keV.
% Note also the presence of a mismatch between the two spectra
%close to 3 keV.  

%%%%%%%%%%%%%%%%%%%%%%%%%%%%%%%%%%%%%%%%%%%%%%%%%%%%%%%%%%5

For the Suzaku/XIS data, we adopt a
0.5-10 keV energy range for the XIS0+XIS2+XIS3 (hereafter XIS023) and
%
%  IMPORTANTE 
%
XIS1 spectra. We exclude the energy interval between 1.7 and 2.4 keV
in the XIS023 and  XIS1, because of systematic features
associated with neutral silicon and neutral gold edges.  We group
the HXD/PIN spectrum to have 25 photons per channel and use the
energy range between 15 and 36 keV. We show in Fig. \ref{backpin} the
HXD/PIN source spectrum and the summed HXD/PIN NXB and CXB spectra
(hereafter NXB+CXB spectrum). The NXB+CXB spectrum dominates the source spectrum at
energies higher than 36 keV.

We analyse the Chandra/MEG  in the
0.5-7 keV  energy range.
Finally, we analyse the JEM-X and IBIS spectrum in the 5-22 and  21-60 keV 
energy band, respectively. The IBIS spectrum is background-dominated above 60 keV.  

\section{Spectral analysis} 

\begin{table*}[ht]
\tiny
  \caption{Best-fit values of the continuum emission}
% title of Table
\label{TAB_Continuum}      % is used to refer this table in the text
%\centering                                      % used for centering table
\begin{center}
\begin{tabular}{l c c c c |c c c c }          % centered columns (4 columns)
                                      % used for centering table
\hline\hline                        % inserts double horizontal lines

%& { {\tt gabs*Comptt}} &
%{\tt cyclabs*Comptt}}\\
      &  \multicolumn{4}{c|}{ MOS12 spectrum in 0.6-7 keV}&
    \multicolumn{4}{c}{ MOS12 spectrum in 1.5-7 keV} \\

     &XMM & Suzaku  & Chandra/MEG & INTEGRAL &XMM & Suzaku  & Chandra/MEG & INTEGRAL \\

\hline  
 \multicolumn{9}{}{ }\\

Parameters    &  \multicolumn{8}{c}{ {\tt     Model: Comptt}} \\

 N$_H$  (10$^{22}$ cm$^{-2}$)&  
 \multicolumn{4}{c|}{$0.126 \pm 0.003$}&
 \multicolumn{4}{c}{$0.121 \pm 0.003$}\\
 
 N$_{H_{pc}}$  (10$^{22}$ cm$^{-2}$)&   
 \multicolumn{4}{c|}{$4.48 \pm 0.08$} &
 \multicolumn{4}{c}{$4.57 \pm 0.08$} \\

$f$ &
 \multicolumn{4}{c|}{$0.633 \pm 0.005$}&
 \multicolumn{4}{c}{$0.638 \pm 0.005$}\\

 E$_{{\rm Edge 1}}$  (keV)&   
 \multicolumn{4}{c|}{$7.20^{+0.02}_{-0.03}$}&
 \multicolumn{4}{c}{$7.20^{+0.02}_{-0.03}$}\\

 $\tau_{{\rm Edge 1}}$  &   
 $0.076 \pm 0.013$ & $0.130 \pm 0.011$& 
$0.076 \pm 0.013$  & $< 0.084$&

 $0.072 \pm 0.013$ & $0.129 \pm 0.011$& 
$0.072 \pm 0.013$  & $< 0.084$\\

 E$_{{\rm Edge 2}}$  (keV)&   
 \multicolumn{4}{c|}{$8.42 \pm 0.05$}&
 \multicolumn{4}{c}{$8.43 \pm 0.05$}\\

 $\tau_{{\rm Edge 2}}$  &   
$0.09 \pm 0.02$ & $0.084 \pm 0.013$&
$0.09 \pm 0.02$ & $0.13^{+0.08}_{-0.10}$&

$0.10 \pm 0.02$ & $0.084 \pm 0.013$&
$0.10 \pm 0.02$ & $0.12^{+0.08}_{-0.10}$\\

 & & & & \\                      

kT$_{0}$ (keV)&
 \multicolumn{4}{c|}{$0.059 \pm 0.006$ }&
 \multicolumn{4}{c}{$0.058 \pm 0.006$ }\\

kT$_{e}$ (keV)&
$3.55\pm 0.11$& $4.32\pm 0.05$&
$3.2^{+0.4}_{-0.3}$& $4.89 \pm 0.08$&

$3.63\pm 0.12$& $4.33\pm 0.05$&
$3.3^{+2.8}_{-0.3}$& $4.89 \pm 0.08$\\

$\tau$ &
$9.60 \pm 0.15  $&$9.09 \pm 0.09  $&
$9.7^{+0.3}_{-0.5}  $&$6.0 \pm 0.2 $&

$9.43 \pm 0.15  $&$9.04 \pm 0.09  $&
$9.6^{+0.4}_{-1.4}  $&$6.0 \pm 0.2 $\\

N$_{CompTT}$ ($10^{-2}$) &
 \multicolumn{4}{c|}{$6.6^{+0.3}_{-0.2} $}&
 \multicolumn{4}{c}{$6.5 \pm 0.3 $}\\

 & & & & \\

$\chi_{red}^2(d.o.f.)$&  
 \multicolumn{4}{c|}{ 1.21(2496)}&
 \multicolumn{4}{c}{ 1.10(2466)}\\

\hline  

  \multicolumn{9}{}{ }\\
                   
   & \multicolumn{8}{c}{ {\tt Model:  gabs*Comptt}} \\

N$_H$  (10$^{22}$ cm$^{-2}$)&  
 \multicolumn{4}{c|}{$0.100 \pm 0.006$}&
 \multicolumn{4}{c}{$0.100 \pm 0.006$}\\
 
 N$_{H_{pc}}$  (10$^{22}$ cm$^{-2}$)&   
 \multicolumn{4}{c|}{$4.54 \pm 0.08$} &
 \multicolumn{4}{c}{$4.61 \pm 0.08$} \\

$f$ &
 \multicolumn{4}{c|}{$0.645 \pm 0.005$}&
 \multicolumn{4}{c}{$0.647 \pm 0.005$}\\

 E$_{{\rm Edge 1}}$  (keV)&   
 \multicolumn{4}{c|}{$7.20^{+0.02}_{-0.04}$}&
\multicolumn{4}{c}{$7.20^{+0.02}_{-0.04}$}\\

 $\tau_{{\rm Edge 1}}$  &   
 $0.075 \pm 0.013$ & $0.125 \pm 0.011$& 
$0.075 \pm 0.013$  & $<0.084$&

 $0.071 \pm 0.013$ & $0.126 \pm 0.011$& 
$0.071 \pm 0.013$  & $<0.084$\\

 E$_{{\rm Edge 2}}$  (keV)&   
 \multicolumn{4}{c|}{$8.43 \pm 0.05$}&
 \multicolumn{4}{c}{$8.43 \pm 0.05$}\\

 $\tau_{{\rm Edge 2}}$  &   
$0.10 \pm 0.02$ & $0.081 \pm 0.013$&
$0.10 \pm 0.02$ & $0.12^{+0.08}_{-0.10}$&

$0.10 \pm 0.02$ & $0.081 \pm 0.013$&
$0.10 \pm 0.02$ & $0.12^{+0.08}_{-0.10}$\\

 E$_{{\rm gabs}}$ (keV)&
 \multicolumn{4}{c|}{$0.73 \pm 0.03$}&
 \multicolumn{4}{c}{$0.72 \pm 0.03$}\\

 $\sigma_{{\rm gabs}}$ (keV)&
 \multicolumn{4}{c|}{$0.14 \pm 0.03$}&
\multicolumn{4}{c}{$0.13 \pm 0.03$}\\
 
 $\tau_{{\rm gabs}}$ &
 \multicolumn{4}{c|}{$0.038^{+0.014}_{-0.009}$  }&
\multicolumn{4}{c}{$0.031^{+0.013}_{-0.009}$  }    \\
 & & & & \\                      

kT$_{0}$ (keV)&
 \multicolumn{4}{c|}{$0.056 \pm 0.010  $  }&
 \multicolumn{4}{c}{$0.052 \pm 0.011  $  }\\

kT$_{e}$ (keV)&
$3.61^{+0.12}_{-0.10}$& $4.33\pm 0.05$&
$3.2 ^{+0.5}_{-0.3}$& $4.89 \pm 0.08$&

$3.67 \pm 0.12$& $4.33\pm 0.05$&
$3.2 ^{+0.6}_{-0.3}$& $4.89 \pm 0.08$\\

$\tau$ &
$9.5 \pm 0.2  $&$9.00 \pm 0.10  $&
$9.6 ^{+0.4}_{-0.5}   $&$6.0 \pm 0.2 $&

$9.3 \pm 0.2  $&$8.98 \pm 0.09  $&
$9.6 ^{+0.4}_{-0.6}   $&$6.0 \pm 0.2 $\\

N$_{CompTT}$ ($10^{-2}$) &
 \multicolumn{4}{c|}{$6.7^{+0.5}_{-0.4}  $}&
 \multicolumn{4}{c}{$6.7 \pm 0.6   $}\\
 & & & & \\                      

$\chi_{red}^2(d.o.f.)$&  
 \multicolumn{4}{c|}{ 1.12(2493)}&
 \multicolumn{4}{c}{ 1.04(2463)}\\

\hline  

              \multicolumn{9}{}{ }\\
      
   & \multicolumn{8}{c}{ {\tt Model: cyclabs*Comptt}} \\
                 
 N$_H$  (10$^{22}$ cm$^{-2}$)&  
 \multicolumn{4}{c|}{$0.100 \pm 0.006$}&
 \multicolumn{4}{c}{$0.104 \pm 0.004$}\\
 
 N$_{H_{pc}}$  (10$^{22}$ cm$^{-2}$)&   
 \multicolumn{4}{c|}{$4.54 \pm 0.08$} &
\multicolumn{4}{c}{$4.75^{+0.16}_{-0.10}$} \\

$f$ &
 \multicolumn{4}{c|}{$0.645 \pm 0.005$}&
 \multicolumn{4}{c}{$0.643 \pm 0.005$}\\

 E$_{{\rm Edge 1}}$  (keV)&   
 \multicolumn{4}{c|}{$7.20^{+0.02}_{-0.04}$}&
 \multicolumn{4}{c}{$7.20^{+0.02}_{-0.04}$}\\

 $\tau_{{\rm Edge 1}}$  &   
 $0.075 \pm 0.013$ & $0.125 \pm 0.011$& 
$0.075 \pm 0.013$  & $<0.084$&

 $0.069 \pm 0.013$ & $0.122 \pm 0.011$& 
$0.069 \pm 0.013$  & $<0.084$\\

 E$_{{\rm Edge 2}}$  (keV)&   
 \multicolumn{4}{c|}{$8.43 \pm 0.05$}&
 \multicolumn{4}{c}{$8.43 \pm 0.05$}\\

 $\tau_{{\rm Edge 2}}$  &   
$0.10 \pm 0.02$ & $0.078 \pm 0.013$&
$0.10 \pm 0.02$ & $0.12^{+0.08}_{-0.10}$&

$0.10 \pm 0.02$ & $0.079 \pm 0.013$&
$0.10 \pm 0.02$ & $0.12^{+0.08}_{-0.10}$\\

 E$_{{\rm cyclabs}}$ (keV)&
 \multicolumn{4}{c|}{$0.69^{+0.03}_{-0.04} $}&
 \multicolumn{4}{c}{$0.68^{+0.04}_{-0.05} $}\\

 W$_{{\rm cyclabs}}$ (keV)&
  \multicolumn{4}{c|}{$0.16 \pm 0.04$}&
  \multicolumn{4}{c}{$0.15 \pm 0.04$}\\

 Dept$_{{\rm cyclabs}}$ &
 \multicolumn{4}{c|}{$0.13 \pm 0.02$ }&
 \multicolumn{4}{c}{$0.12 \pm 0.02$ }\\
 & & & & \\                      

kT$_{0}$ (keV)&
 \multicolumn{4}{c|}{$0.055 \pm 0.010 $  }&
 \multicolumn{4}{c}{$0.051 \pm 0.011 $  }\\

kT$_{e}$ (keV)&
$3.64\pm 0.13$& $4.34\pm 0.05$&
$3.2^{+0.5}_{-0.3}$& $4.89 \pm 0.08$&

$3.70^{+0.13}_{-0.11}$& $4.34\pm 0.05$&
$3.2^{+2.8}_{-0.3}$& $4.89 \pm 0.08$\\

$\tau$ &
$9.4 \pm 0.2  $&$8.91 \pm 0.10  $&
$9.5^{+0.4}_{-0.5}   $&$6.0 \pm 0.2 $&

$9.2 \pm 0.2  $&$8.91^{+0.10}_{-0.18}   $&
$9.6^{+0.4}_{-1.5}   $&$6.0 \pm 0.2 $\\

N$_{CompTT}$ ($10^{-2}$) &
 \multicolumn{4}{c|}{$6.8 \pm 0.5 $}&
 \multicolumn{4}{c}{$6.9 \pm 0.6 $}\\
 & & & & \\                      

$\chi_{red}^2(d.o.f.)$&  
 \multicolumn{4}{c|}{ 1.12(2493)}&
 \multicolumn{4}{c}{ 1.04(2463)}\\

\hline                                             %inserts single line
\end{tabular}
\end{center}

{\small \sc Note} \footnotesize---Uncertainties are at the  90\%
confidence level for a single parameter.  
 The  $\chi_{red}^2(d.o.f.)$ 
 values are obtained taking 
the emission  lines shown in Table \ref{TAB_Line}   into account. 
\end{table*}
\begin{table*}[ht]
  \caption{Best-fit values of the emission lines }
% title of Table
\label{TAB_Line}      % is used to refer this table in the text
%\centering      
\tiny                          % used for centering table
\begin{center}
\begin{tabular}{l c c c|c c c }          % centered columns (4 columns)
\hline\hline                        % inserts double horizontal lines
      &  \multicolumn{3}{c|}{ MOS12 spectrum in 0.6-7 keV}&
    \multicolumn{3}{c}{ MOS12 spectrum in 1.5-7 keV} \\

% &  \multicolumn{3}{c|}{ {\tt gabs*Comptt}} & \multicolumn{3}{c}{ 
%{\tt cyclabs*Comptt}}\\

 Line  &  E (keV) & $\sigma$ (eV) & I $(\times 10^{-4})$&  E (keV) & $\sigma$ (eV) & I $(\times 10^{-4})$\\

\hline  
      \multicolumn{7}{}{ }\\

            &  \multicolumn{6}{c}{ {\tt Model: Comptt}}\\ 

                             %inserts single line
  
\ion{N}{vii} 
&   0.5 (fixed) &  1 (fixed) &  $2.8    \pm 0.9$ &
    0.5 (fixed) &  1 (fixed) &  $2.5    \pm 0.9$ \\

\ion{O}{vii} (i)
&   0.5687 (fixed) &  1.5 (fixed)&  $19 \pm 2$ &
    0.5687 (fixed) &  1.5 (fixed)&  $17 \pm 2$ \\

\ion{O}{viii} & 
               0.6536 (fixed) & $ 1$ (fixed) & $1.8 \pm 0.5$ &
               0.6536 (fixed) & $ 1$ (fixed) & $1.6 \pm 0.5$ \\

\ion{Ne}{ix} (i) & 
                   0.915 (fixed) & $ 3$ (fixed)  & $2.5 \pm 0.3$ &
                   0.915 (fixed) & $ 3$ (fixed)  & $2.6^{+0.9}_{-0.3}$ \\

\ion{Ne}{x}  & 
                   1.022 (fixed) & $ 3$ (fixed)  & $0.9 \pm 0.2$ &
                   1.022 (fixed) & $ 3$ (fixed)  & $0.9^{+0.5}_{-0.2}$ \\

\ion{Mg}{xi} (i)  & 
                   1.3434 (fixed) & $ 3$ (fixed)  & $0.72 \pm 0.14$ &
                   1.3434 (fixed) & $ 3$ (fixed)  & $0.69^{+0.48}_{-0.13}$ \\

\ion{Mg}{xii}   & 
                   1.4726 (fixed)  & $ 3$ (fixed)  & $0.40 \pm 0.06$ &
                   1.4726 (fixed)  & $ 3$ (fixed)  & $0.34^{+0.27}_{-0.09}$ \\

\ion{Si}{xiv}   & 
                   2.005  (fixed)& $ 3$ (fixed)  & $0.35 \pm 0.10$ &
                   2.005  (fixed)& $ 3$ (fixed)  & $0.35^{+0.27}_{-0.08}$ \\

\ion{Fe}{i}    & 
                 $6.408 \pm 0.005$ & $20 \pm 15$   & $2.31 \pm 0.15$&
                 $6.408 \pm 0.005$ & $22 \pm 15$   & $2.30 \pm 0.15$\\

\ion{Fe}{xxvi}    & 
                    $6.98 \pm 0.03$ & $50$ (fixed)  & $0.59 \pm 0.13$&
                    $6.98 \pm 0.03$ & $50$ (fixed)  & $0.60 \pm 0.14$\\
\hline                    
       \multicolumn{7}{}{ }\\
            &  \multicolumn{6}{c}{ {\tt Model: gabs*Comptt}}\\ 
                             %inserts single line
   
\ion{N}{vii} &
                 0.5 (fixed) &   1 (fixed) &  $1.7   \pm 0.8$ &
                 0.5 (fixed) &   1 (fixed) &  $1.7   \pm 0.8$ \\

\ion{O}{vii} (i) &  
                     0.5687 (fixed) & 1.5 (fixed)&  $14.8 \pm 1.5$ &
                     0.5687 (fixed) & 1.5 (fixed)&  $14.6 \pm 1.5$ \\

\ion{O}{viii} & 
               0.6536 (fixed) & $ 1$ (fixed) & $1.8 \pm 0.5$ &
               0.6536 (fixed) & $ 1$ (fixed) & $1.8 \pm 0.5$ \\

\ion{Ne}{ix} (i) & 
                   0.915 (fixed) & $ 3$ (fixed)  & $2.6 \pm 0.4$ &
                   0.915 (fixed) & $ 3$ (fixed)  & $2.6^{+1.1}_{-0.4}$ \\

\ion{Ne}{x}  & 
                   1.022 (fixed) & $ 3$ (fixed)  & $0.7\pm 0.2$ &
                   1.022 (fixed) & $ 3$ (fixed)  & $0.8^{+0.4}_{-0.2}$ \\

\ion{Mg}{xi} (i)  & 
                   1.3434 (fixed) & $ 3$ (fixed)  & $0.68 \pm 0.14$ &
                   1.3434 (fixed) & $ 3$ (fixed)  & $0.66^{+0.36}_{-0.13}$ \\

\ion{Mg}{xii}   & 
                   1.4726 (fixed)  & $ 3$ (fixed)  & $0.39 \pm 0.06$ &
                   1.4726 (fixed)  & $ 3$ (fixed)  & $0.34^{+0.21}_{-0.10}$ \\

\ion{Si}{xiv}   & 
                   2.005  (fixed)& $ 3$ (fixed)  & $0.40 \pm 0.05$ &
                   2.005  (fixed)& $ 3$ (fixed)  & $0.36^{+0.25}_{-0.08}$ \\

\ion{Fe}{i}    & 
                 $6.408 \pm 0.003$ & $23 \pm 10$   & $2.4 \pm 0.2$&
                 $6.408 \pm 0.005$ & $24 \pm 15$   & $2.3 \pm 0.2$\\

\ion{Fe}{xxvi}    & 
                    $6.98 \pm 0.02$ & $50$ (fixed)  & $0.62 \pm 0.13$&
                    $6.98 \pm 0.03$ & $50$ (fixed)  & $0.62 \pm 0.14$\\
                    
\hline                    
       \multicolumn{7}{}{ }\\
    
            &  \multicolumn{6}{c}{ {\tt Model: cyclabs*Comptt}}\\ 
                             %inserts single line
   
\ion{N}{vii} &
                0.5 (fixed) &  1 (fixed) &  $1.6   \pm 0.8$ &
                0.5 (fixed) &  1 (fixed) &  $1.6   \pm 0.8$ \\

\ion{O}{vii} (i) &
                  0.5687 (fixed) &  1.5 (fixed)&  $14.8 \pm 1.5$ &
                  0.5687 (fixed) &  1.5 (fixed)&  $14.6 \pm 1.5$ \\

\ion{O}{viii} & 
               0.6536 (fixed) & $ 1$ (fixed) & $1.8 \pm 0.5$ &
               0.6536 (fixed) & $ 1$ (fixed) & $1.8 \pm 0.5$ \\

\ion{Ne}{ix} (i) & 
                   0.915 (fixed) & $ 3$ (fixed)  & $2.6 \pm 0.4$ &
                   0.915 (fixed) & $ 3$ (fixed)  & $2.6^{+1.0}_{-0.4}$ \\

\ion{Ne}{x}  & 
                   1.022 (fixed) & $ 3$ (fixed)  & $0.8\pm 0.2$ &
                   1.022 (fixed) & $ 3$ (fixed)  & $0.8^{+0.5}_{-0.2}$ \\

\ion{Mg}{xi} (i)  & 
                   1.3434 (fixed) & $ 3$ (fixed)  & $0.64 \pm 0.14$ &
                   1.3434 (fixed) & $ 3$ (fixed)  & $0.63^{+0.37}_{-0.13}$ \\

\ion{Mg}{xii}   & 
                   1.4726 (fixed)  & $ 3$ (fixed)  & $0.32 \pm 0.10$ &
                   1.4726 (fixed)  & $ 3$ (fixed)  & $0.32^{+0.21}_{-0.10}$ \\

\ion{Si}{xiv}   & 
                   2.005  (fixed)& $ 3$ (fixed)  & $0.37 \pm 0.09$ &
                   2.005  (fixed)& $ 3$ (fixed)  & $0.37^{+0.27}_{-0.08}$ \\

\ion{Fe}{i}    & 
                 $6.408 \pm 0.005$ & $26 \pm 15$   & $2.4 \pm 0.2 $&
                 $6.408 \pm 0.005$ & $26 \pm 15$   & $2.4 \pm 0.2 $\\

\ion{Fe}{xxvi}    & 
                    $6.98 \pm 0.03$ & $50$ (fixed)  & $0.64 \pm 0.14$&
                    $6.98 \pm 0.03$ & $50$ (fixed)  & $0.65 \pm 0.14$\\

\hline                                             %inserts single line
\end{tabular}
\end{center}
{\small \sc Note} \footnotesize---Uncertainties are at the  90\%
confidence level  for a single parameter. The line intensities are in units 
of photons cm$^{-2}$ s$^{-1}$.    
\end{table*}

We simultaneously fitted the XMM-Newton, Suzaku, Chandra, and INTEGRAL
  spectra.  We add ed a systematic
  error of 1\% to take into account that the observations are not simultaneous.
Initially, we fitted the spectra using the 0.6-7 keV energy range for the MOS12.

 We fitted the spectra using
XSPEC  version 12.8.1 \citep[see][]{Arnaud_96}.  Initially, to fit the continuum emission, we
adopted the model used by \cite{iaria_2013}. It is {\tt
  Ed*phabs*(f*cabs*phabs*(LN+CompTT)+(1-f)*(LN+CompTT))}, which is a
Comptonised component ({\tt CompTT} in XSPEC) absorbed by neutral
interstellar matter (the first {\tt phabs} component) and partially
absorbed by local neutral matter (the second {\tt phabs} component).
We used the abundances provided by \cite{aspl} and the photoelectric
cross-section given by \cite{verner_96}. We took the Thomson
scattering of the local neutral matter into account by adding the {\tt
  cabs} component and imposed that the equivalent hydrogen column
density of the {\tt cabs} component is the same as the local neutral
matter.  The constant {\tt f} gives the percentage of emitting region
occulted by the local neutral matter.  Finally, {\tt LN} and {\tt Ed}
in the model indicate all the Gaussian components added to the model
to fit the several emission lines observed in the spectrum and the
added absorption edges, respectively.

Since the XMM-Newton, Suzaku, Chandra, and INTEGRAL  observations are not
simultaneous we left the values of 
 the electron temperature and of the optical depth of the
{\tt CompTT} component free to vary independently. The depths of the absorption edges added above 7 keV 
were free to vary independently for the XMM-Newton, Suzaku, and INTEGRAL spectra, whilst
they were tied in the Chandra spectrum to the values of  the XMM-Newton spectrum
because the Chandra spectrum extends up to 7 keV. 
 \begin{figure}[ht]
 \resizebox{\hsize}{!}{\includegraphics{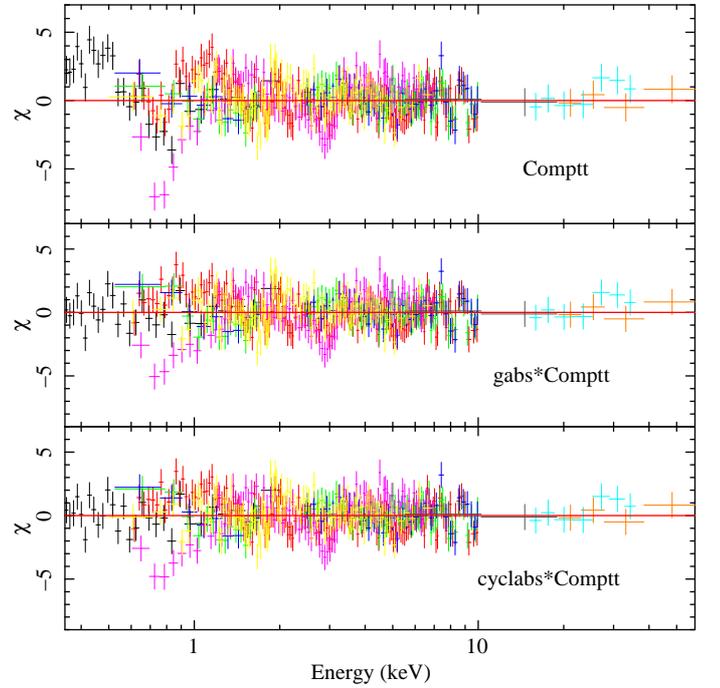}}
\caption{Residuals with respect to the best-fit models shown in
  Tables \ref{TAB_Continuum} and \ref{TAB_Line}.  
 The RGS12, EPIC-pn, XIS023, XIS1, MOS12, HXD/PIN, MEG, JEM-X, and IBIS
  spectra are shown in black, red, green, blue, magenta, light-blue, yellow, 
 grey,  and orange, respectively. The data are graphically rebinned.
  From top to bottom, the residuals with respect to the continuum consist of: 1) 
  a {\tt Comptt} partially absorbed by local
  neutral matter (large residuals are evident at 0.7 keV); 2) {\tt
    gabs*Comptt} with the energy of the {\tt gabs} component close to
  0.73 keV; 3) {\tt cyclabs*Comptt}. The MOS12 spectrum covers the 0.6-7 keV energy band.}
\label{Fig1}
\end{figure}
 \begin{figure}[ht]
 \resizebox{\hsize}{!}{\includegraphics{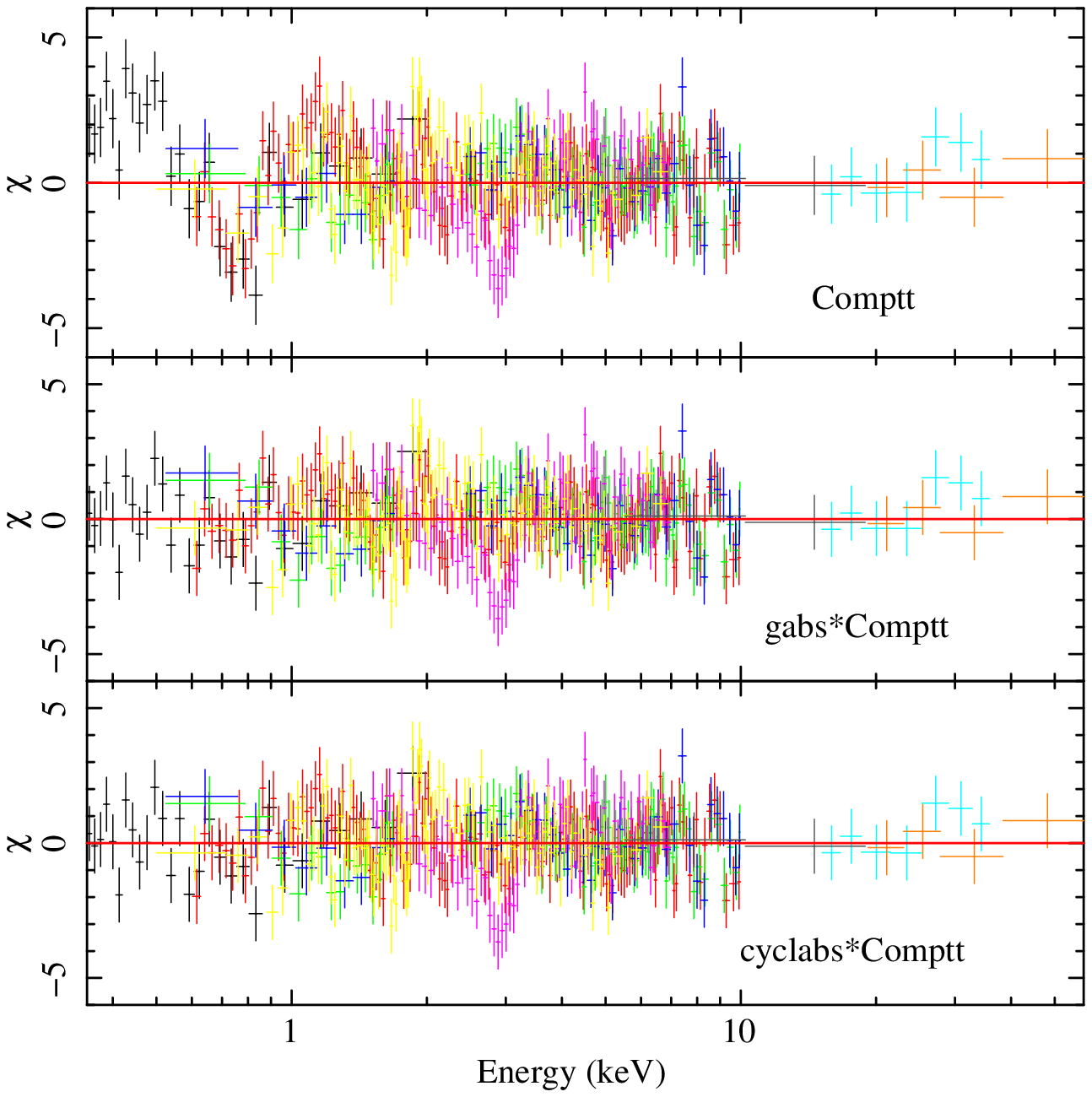}}
\caption{Residuals with respect to the best-fit models shown in
  Tables \ref{TAB_Continuum} and \ref{TAB_Line}.  The colours 
are defined as in other figures. The data are graphically rebinned.
  From top to bottom, the residuals  with respect to the
  continuum consist of: 1) a {\tt Comptt} partially absorbed by local
  neutral matter  (large residuals are evident at 0.7 keV); 2) {\tt
    gabs*Comptt} with the energy of the {\tt gabs} component close to
  0.72 keV; 3) {\tt cyclabs*Comptt}. 
The MOS12 spectrum covers the 1.5-7 keV energy band.}
\label{Fig1b}
\end{figure}

Several emission lines are detected and identified with \ion{N}{vii},
\ion{O}{vii} intercombination line, \ion{O}{viii}, \ion{Ne}{ix}
intercombination line, \ion{Ne}{x}, \ion{Mg}{xi} intercombination
line, \ion{Mg}{xii}, \ion{Si}{xiv}, \ion{Fe}{i,} and \ion{Fe}{xxvi}. The emission lines are fitted with Gaussian components.
We fixed the energies and widths of the emission lines below 6 keV at
their best-fit values because their analysis is not the aim of this
work; a detailed analysis of these lines is reported by
\cite{iaria_2013}.  Finally, we added two absorption edges at 7.2 and
8.4 keV.  Fitting the spectra we obtain a $\chi^2({\rm {d.o.f.}})$ of
 3024(2496) and large residuals between 0.35 and 1 keV are
visible. We show the residuals in Fig. \ref{Fig1} (top panel),
 the best-fit
values of the continuum  emission and absorption edges in
Table \ref{TAB_Continuum}, and the best-fit parameters associated with the
emission lines in
Table \ref{TAB_Line}. 

 To fit the large residuals between 0.4 and 1 keV, we added a
  Gaussian absorption line ({\tt gabs} in XSPEC) that is
  a multiplicative component. The component {\tt gabs} is defined by
  three parameters, and its functional form is 
$$M(E) =
\exp[-(\tau/\sqrt{2 \pi} \sigma) \exp(-((E-E_0)/4\sigma))^2],$$ where
$\tau$, $\sigma$, and $E_0$ are the line depth, the line width in keV,
and the line energy in keV, respectively.    We interpret this component as a
CRSF in the spectrum.  The model becomes {\tt
  Ed*phabs*(f*cabs*phabs*(LN+gC)+(1-f)*(LN+gC))}, where {\tt gC} is
{\tt gabs*CompTT} and {\tt Ed} takes the absorption edges into
account.  The addition of the {\tt gabs} component improves the fit.
We obtain a $\chi^2({\rm {d.o.f.}})$ of  2798(2493) with a
$\Delta\chi^2 $ of  226 and a F-statistics value of 67.1. The
residuals are shown in Fig. \ref{Fig1} (the second panel from the top).
The best-fit parameters are shown in Tables \ref{TAB_Continuum}
 and \ref{TAB_Line}.  
%We show the
%data/model ratio of the best-fit model excluding the {\tt gabs}
%component in Fig. \ref{Fig2}.
 \begin{figure}[ht]
 \resizebox{\hsize}{!}{\includegraphics{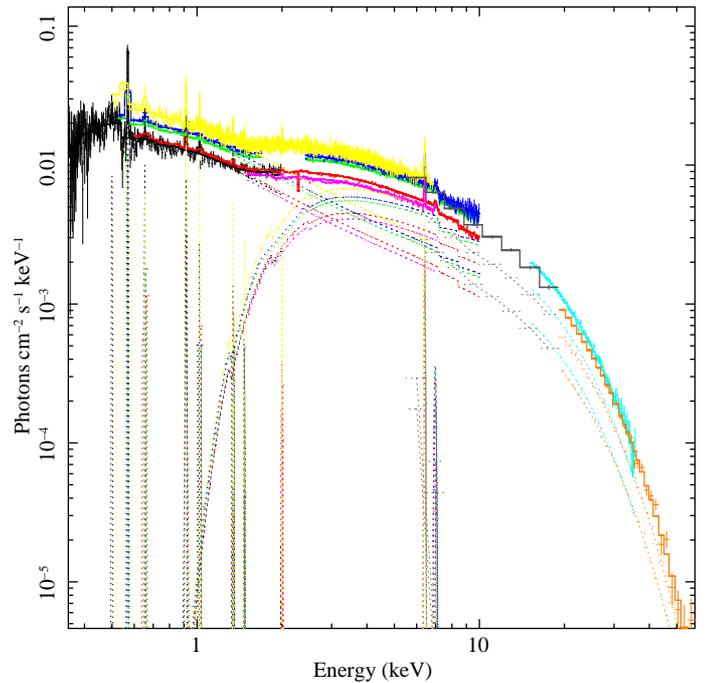}}
\caption{Unfolded spectra  relative to the model including the {\tt gabs} component. 
The MOS12 spectrum ranges between 1.5 and 7 keV. Colours as above.  }
\label{Fig1c}
\end{figure}

 \begin{figure}[ht]
\resizebox{\hsize}{!}{\includegraphics{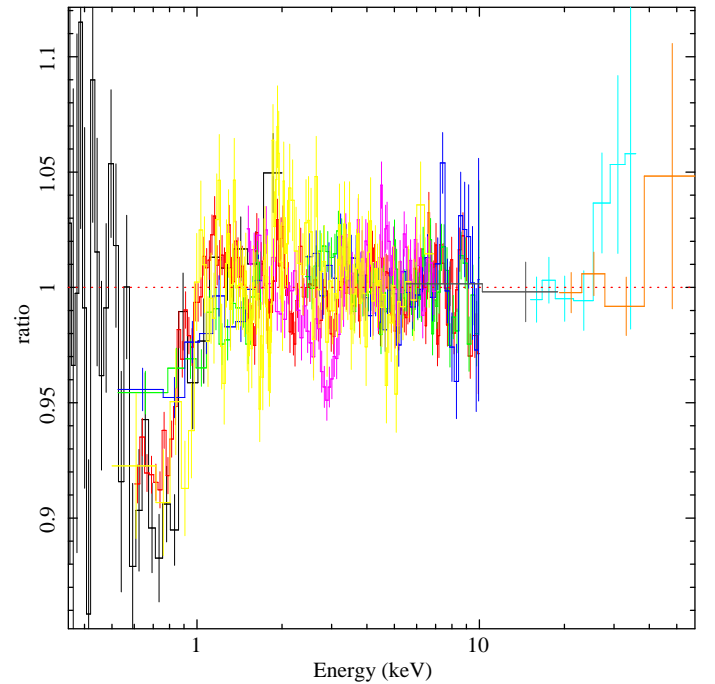}}
\caption{Data/model ratio  with respect to the best-fit model shown in
  Tables \ref{TAB_Continuum} and \ref{TAB_Line},  but excluding the {\tt
    gabs} component. The data are graphically rebinned.  Colours as above. The MOS12 spectrum ranges between 1.5
  and 7 keV. }
\label{Fig2}
\end{figure}

 We also fitted the residuals between 0.4 and 1 keV using, instead of
  a Gaussian absorption line, an absorption line with a Lorentzian
  shape \citep[][]{mihara_90}. This multiplicative component ({\tt
    cyclabs} in XSPEC) is defined by three parameters, and its
  functional form is $$M(E) = \exp[- \tau\; (\sigma E /E_0)^2
  /[(E-E_0)^2+\sigma^2] ],$$ where $\tau$, $\sigma$, and $E_0$ are the
  line depth, the line width in keV, and the line energy in keV,
  respectively.  We  fixed the second harmonic
   depth to zero in the model.  In this case the adopted model is {\tt
  Ed*phabs*(f*cabs*phabs*(LN+cC)+(1-f)*(LN+cC))}, where {\tt cC} is
{\tt cyclabs*CompTT}.  The addition of the {\tt cyclabs} component
instead of the {\tt gabs} component gives an equivalent fit with a
$\chi^2({\rm {d.o.f.}})$ of 2795(2493). The best-fit values 
 are shown in
Tables  \ref{TAB_Continuum} and
\ref{TAB_Line}. The
residuals are shown in Fig. \ref{Fig1} (bottom  panel).

The residuals in Fig. \ref{Fig1} show that the RGS12, XIS1, XIS023,
MEG, and EPIC-pn spectra are in good agreement below 1.5 keV, unlike in the MOS12 spectrum. For this reason we repeated the analysis
described above excluding the 0.6-1.5 keV energy range in the MOS12
spectrum.  Fitting the spectra using the initial model, we find a
$\chi^2({\rm {d.o.f.}})$ of 2713(2466) and large residuals between 0.35
and 1 keV are visible (see Fig. \ref{Fig1b}, top panel).  Using the
model that includes the {\tt gabs} component, we obtain a $\chi^2({\rm
  {d.o.f.}})$ of 2565(2463) with a $\Delta\chi^2 $ of 148 and a
F-statistics value of 47.4.  The residuals are shown in
Fig. \ref{Fig1b} (middle panel).  Using the model including the {\tt
  cyclabs} component, we obtain a $\chi^2({\rm {d.o.f.}})$ of
2565(2463).  The residuals are shown in Fig. \ref{Fig1b} (bottom
panel). The unfolded spectra  relative to the model including the {\tt gabs}
component is shown in Fig. \ref{Fig1c}.  We show the data/model ratio with respect to
the best-fit model, but excluding the {\tt gabs} component in
Fig. \ref{Fig2}.

Furthermore,
using the initial model with the MOS12 spectrum between  1.5-7 keV, we   
look for the presence of  CRSF at high energies, 
as suggested by \cite{sasano_13},
exploiting the availability of  the JEM-X and IBIS spectra. 
Adding to the model a {\tt cyclabs} component with width and centroid
fixed  to 5 keV and 33 keV \citep[see][]{sasano_13}, respectively, we find an
upper limit on the depth of 0.10 at 99.7\% confidence level (3$\sigma$). 
This value is not consistent with the $0.4^{+0.2}_{-0.1}$ obtained by
\cite{sasano_13}.

 We find that the energy and width of the CRSF at 0.7 keV are
  consistent for the Lorentzian and the Gaussian shapes. 
The depth values of the absorption edges at 7.2 keV are compatible 
for the XMM-Newton and  INTEGRAL spectra, while they higher in the Suzaku spectrum.
Finally, the optical depth of the Comptonised component assumes similar values 
close to nine in the XMM-Newton, Chandra, and Suzaku spectra, while it is $6.0 \pm 0.2$
in the INTEGRAL spectrum.  

Finally, we note that including  MOS12 spectrum
between 0.6 and 1.5 keV  only marginally affects the fit results. 
In the following we use the best-fit values obtained when excluding the
0.6-1.5 keV energy range of MOS12.

\section{Discussion} 

We used three non-simultaneous pointed X-ray observations of
X1822-371: a XMM-Newton observation (using RGS, MOS, and EPIC-pn spectra), a
Suzaku observation (using XIS and HXD/PIN spectra), and finally, a
Chandra observation (using the first-order MEG spectrum). The spectral
analysis of the XMM and Chandra data sets have already been discussed in
\cite{iaria_2013}. The same XMM/EPIC-pn data were also analysed by
\cite{somero_12}, who report a different interpretation of the X-ray
spectrum.  Moreover, we used all the available INTEGRAL/JEM-X and
INTEGRAL/ISGRI observations of X1822-371 and extracted the
corresponding spectra to  confirm or disprove a claimed CRSF at
33 keV in the Suzaku/PIN spectrum \citep[see][]{sasano_13}.  

We 
adopted the same model as proposed by those authors to fit the continuum
emission. It consists of a Comptonised component {\tt Comptt}
absorbed by interstellar matter and partially absorbed by local
neutral matter. The effect of Thomson scattering on the local cold
absorber is taken into account by adding the {\tt cabs} component with an
equivalent hydrogen column density imposed to be the same as the one
associated with the local neutral matter. \cite{iaria_2013} found
residuals in the EPIC-pn data between 0.6 and 0.8 keV and added a
black-body component with temperature fixed to 0.06 keV, thereby improving the fit
significantly.
 
In this work we give a different interpretation of the residuals in
the EPIC-pn data between 0.4 and 1 keV, modelling them with the
addition of a CRSF close to 0.7 keV.  The broad residuals are fitted
using the {\tt gabs} or the {\tt cyclabs}  model components.  The
addition of this component  significantly improves
the fit, and we obtain statistically equivalent fits using  either
the {\tt gabs}   or the {\tt cyclabs} component.

The interpretation of the residuals as a CRSF in the spectrum allows us
to refine the scenario proposed by \cite{iaria_2013} for X1822-371.
The authors suggested that the Comptonised component originates in the
inner region of the system,   it is not directly observable because of the
large inclination angle of the system, and only 1\% of its flux
arrives to the observer because  of scattering  by an extended
optically thin corona with optical depth $\sim 0.01$. We now suggest
that the Comptonised component could be produced in the accretion
column onto the NS magnetic caps.

%{\bf The spin-phase resolved analysis of the EPIC-pn spectra show 
%  that the spectral shape does not change along the spin period and it
%  is reasonable considering that we observe only the 1\% of the flux
%  scattered from an optically thin extended corona.  }

Below we discuss our results showing that a CRSF at 0.7 keV 
agrees with the non-conservative mass transfer scenario proposed
for X1822-371 in the past three years by several authors \citep[see,
for example,][]{iaria_2013,iaria_2011,Burderi_2010,Bayless2010} and
allows the NS mass of the binary system that is
between 1.61 and 2.32 M$_{\odot}$  to be further constrained\citep{munoz_2005}.
\cite{Burderi_2010}, \cite{Bayless2010}, and \cite{iaria_2011} found a
large orbital period derivative of X1822-371, and \cite{Burderi_2010}
show that this indicates that X1822-371 accretes at the Eddington
limit and that the rest of the mass transferred by the companion star is
expelled from the system.  

Adopting the values for  $P_s$ and $\dot{P_s}$ derived
in Sect. \ref{timing} 
and assuming that X1822-371 accretes at its Eddington limit,  
we show  that the  CRSF energy obtained by our fits is consistent
with a scenario in which the NS in   X1822-371 is spinning up. 
Since $\dot{P_s}$ is negative, the corotation radius $r_c$,
the radius at which the accretion disc has the same angular velocity 
as the NS, has to be
larger than the magnetospheric radius $r_m$,
the radius at which the  magnetic pressure of the 
 NS B-field equals  the ram pressure of the accreting matter. 
The corotation radius can be expressed as 
$r_c=(GM_{NS}/4\pi^2)^{1/3} P_s^{2/3}$, where $M_{NS}$ is the NS mass
and $P_s$ is the NS spin period.  Assuming a NS mass of 1.4
M$_{\odot}$, we obtain $r_c \simeq 1,180$ km; for a NS
mass of 2 M$_{\odot}$, the corotation radius is $r_c \simeq
1,300$ km.  The magnetospheric radius is given by the relation $r_m =
\phi r_A$, where $r_A$ is the Alfven radius and $\phi$ a constant
close to 0.5 \citep[see][]{ghosh_91}. We estimate $r_A$ using the
Eq. 2 in \cite{burderi_98}:
\begin{equation}
\label{alven}
r_A = 4.3 \times 10^3 \mu_{30}^{4/7} R_6^{-2/7} L_{37}^{-2/7} \epsilon^{2/7} m^{1/7} {\rm km},
\end{equation}
where $ \mu_{30}$ is the magnetic moment in units of $10^{30}$ G
cm$^3$, $R_6$ the NS radius in units of $10^6$ cm, $ L_{37}$
the luminosity of the system in units of $10^{37}$ erg/s, $m$ the
NS mass in units of solar masses, and finally, $\epsilon$
is the ratio between the luminosity and the total gravitational
potential energy released per second by the accreting matter.

Adopting the best-fit value of the  CRSF energy, $E_{kev}$,  obtained {\bf from}
the {\tt gabs} component, and assuming that this is produced at (or very close to)
the NS surface, 
 we estimate the NS B-field  
using the relation $E_{kev} = (1+z)^{-1}\;11.6 B_{12}$, where
$$
 (1+z)^{-1} =\left( 1-\frac{2GM_{NS}}{R_{NS}c^2}\right)^{1/2}
$$
and $B_{12}$ is the NS B-field in units of 
$10^{12}$ G. We arrange the term $ (1+z)^{-1}$ in terms of $m$ and $R_6$ and 
obtain
\begin{equation}
\label{B_field}
B_{12} = \frac{E_{kev}}{11.6}\left(1-0.295 \frac{m}{R_6}\right)^{-1/2} {\rm G}.
\end{equation}
We assume that the intrinsic luminosity of X1822-371 is the Eddington
luminosity, $L=1.26 \times 10^{38} (M_{NS}/M_{\odot})$ erg/s, which we
rewrite as $L_{37} = 12.6 \;m$, which is the Eddington luminosity in
units of $10^{37}$ erg/s. Substituting this expression 
of luminosity and the expression of $B_{12}$ of Eq. \ref{B_field}
 into
Eq. \ref{alven} and assuming a NS radius of 10 km, we obtain
$$
r_A = 4.3 \times 10^2 \left(1-0.295 m\right)^{-2/7}  m^{-1/7} {\rm km}.
$$
For NS masses of 1.4 and 2 M$_{\odot}$, we obtain $r_A\simeq 490$ and $\simeq 510$
km, respectively. Since $r_m=\phi r_A$ with $\phi \simeq 0.5$ then 
for a NS mass of 1.4 and 2 M$_{\odot}$, we obtain $r_m\simeq 245$ and $\simeq 255$
km, respectively.  This means that, for a NS mass  between 1.4 and 2  M$_{\odot}$, 
$r_m$ is always smaller than $r_c$  by  a 
factor five. This implies that the accreting matter gives specific angular
momentum to the NS, which increases its angular velocity and spins it up. 

Next we adopt the set of relations shown by \cite{Ghosh_lamb}
(Eqs. from 15 to 18) to establish a relation for the derivative of the spin
period, the spin period, the luminosity, and the NS mass.  This set
of equations is valid for the fastness parameter $\omega_s
=\Omega_s/\Omega_K(r_o)< \omega_{max} \simeq 0.95$, where $\Omega_s$
is the NS angular velocity, $\Omega_K(r_o)$ is the Keplerian angular
velocity at $r_0$, and $r_0$ is the radius that separates the
boundary layer from the outer transition zone \citep[see discussion
in][]{Ghosh_lamb}.  Using Eq. \ref{B_field} and assuming that the
NS accretes at the Eddington limit ($L_{37} = 12.6 \;m$), the
parameter $\omega_s$ of the Eq. 16 in \cite{Ghosh_lamb} becomes
\begin{equation}
\label{omegas}
\omega_s \simeq 0.456 \left(\frac{E_{keV}}{11.6}\right)^{6/7} 
\left(1-0.295 \frac{m}{R_6} \right)^{-3/7} R_6^{15/7} m^{-5/7} P_s^{-1}. 
\end{equation}
Using the value of $E_{keV}$ shown in Table \ref{TAB_Continuum}, the
spin period value shown in Sect. \ref{timing}, and finally, imposing
that $R_6 =1,$ we find that $\omega_s$ is between 0.063 and 0.083 for
$m$ between 1 and 3 M$_{\odot}$. This implies that the values of NS
B-field and luminosity satisfy the spin-up condition.

Using Eq. 15 of \cite{Ghosh_lamb}, we constrained the NS mass. 
We adopted the expression of the NS moment of inertia 
given by  \cite{lattimer_05} in Eq. 16 of their work. The expression  is valid for many
NS equation of states and for a NS mass higher than 1 M$_{\odot}$.  
 Rewriting the expression in terms of $m$ and $R_6$
we obtain
 \begin{equation}
\label{Inertia}
I_{45} \simeq (0.471 \pm 0.016) \; m R_6^2 \left(1+ 0.42 \frac{m}{R_6}
+0.009 \frac{m^4}{R_6^4}\right),
\end{equation}
where $I_{45}$ is the NS moment of inertia in units of $10^{45}$ g cm$^{2}$.
The Eq. 15  of \cite{Ghosh_lamb} can be rewritten as
 \begin{equation}
\label{Pdot}
-\dot{P}_{-12} \simeq 29.51\; R_6^{-2/7} m^{-4/7} B_{12}^{2/7} \delta^{-1}  P_s^{2} n(\omega_s),
\end{equation}
where $\dot{P}_{-12}$ is the spin period derivative in units of
$10^{-12}$ s/s, $\delta$ is the term in parenthesis in
Eq. \ref{Inertia}, $B_{12}$ is given by  Eq. \ref{B_field}, and
finally, the function $n(\omega_s)$ in its useful approximate
expression is
$$
n(\omega_s) \sim 1.39 \{1-\omega_s[4.03(1-\omega_s)^{0.173}-0.878]\}
(1-\omega_s)^{-1}
$$
\citep[see Eq. 10 in][]{Ghosh_lamb}.

Initially, we impose that $R_6=1$ and find how  $\dot{P}_{-12}$ changes 
as function of $m$. 
 \begin{figure}[ht]
\resizebox{\hsize}{!}{\includegraphics[angle=-90]{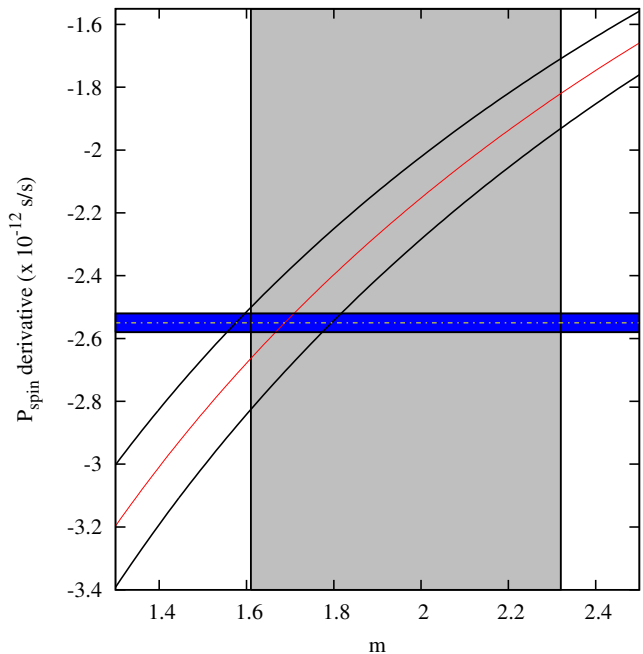}}
\caption{Spin period derivative, $\dot{P}_{-12}$, vs. $m$ for a
  NS radius of 10 km (red curve); the black curves  represent  the upper
  and lower limit values of the spin derivative. The grey box
  indicates the mass range allowed for X1822-371 according to
  \cite{munoz_2005}. The horizontal blue strip indicates the value of
  $\dot{P}_{-12}$  as derived in Sect. \ref{timing}.}
\label{Fig3}
\end{figure}
The error associated with $\dot{P}_{-12}$ is mainly due to the term
$n(\omega_s)$, that has an accuracy of 5\% \citep[see][]{Ghosh_lamb};
we take also the errors associated with $I$ and $E_{keV}$ into
account.  We adopt the 1$\sigma$ error for $E_{keV}$.  We show the
values of $\dot{P}_{-12}$ vs $m$ in Fig.  \ref{Fig3}. For a range of $m$ between 1.3 and 2.5, the
$\dot{P}_{-12}$ changes from -3.4 up to -1.6.  The grey box in Fig.
\ref{Fig3} limits the allowed NS mass between 1.61 and 2.32
M$_{\odot}$, according to \cite{munoz_2005}.  Furthermore, the value
of the spin period derivative, $\dot{P}_{-12} = -2.55 \pm 0.03$ s/s
that we obtained in Sect. \ref{timing}, is shown, as are the
uncertainties associated with $\dot{P}_{-12}$.  We find a NS mass of
$1.69 \pm 0.13$ M$_{\odot}$ assuming  $R_6=1$, which is inside the
range suggested by \cite{munoz_2005}.  Consequently, using the
Eq. \ref{B_field}, we find that the NS B-field is $(8.8 \pm 0.3)
\times 10^{10}$ G, a value that is very similar to the one suggested
by \cite{Jonker_2001} assuming a luminosity of $\sim 10^{38}$ erg/s
for X1822-371.

Using the mass function $(2.03 \pm 0.03) \times 10^{-2}$ M$_{\odot}$
and an inclination angle of X1822-371 of $82\fdg5$
\citep[see][]{Jonker_2003}, we infer  the companion star mass,
$M_c = 0.46 \pm 0.02$ M$_{\odot}$, which is  close to the value
of 0.5 M$_{\odot}$ suggested by \cite{munoz_2005}.   Using only
  optical observations, \cite{somero_12} estimate that the mass
ratio of X1822-371 is $q=M_c/M_{NS} = 0.28$. Using this relation, we
find that the companion star mass is $M_c = 0.47 \pm 0.04$
M$_{\odot}$ for a NS mass of $1.69 \pm 0.13$ M$_{\odot}$, and that value
is compatible with the one inferred by us using the mass function.

\begin{table}[ht]
  \caption{Values of  R$_{NS}$,  M$_{NS}$,  B, and  M$_c$
obtained from the cyclotron line energy found adopting  the {\tt gabs} component in the fit.}
% title of Table
\label{TAB_M_R}      % is used to refer this table in the text
%\centering                                      % used for centering table
\begin{center}
\begin{tabular}{l c  c c}          % centered columns (4 columns)
                                      % used for centering table
\hline\hline                        % inserts double horizontal lines
 R$_{NS}$ & M$_{NS}$ & B & M$_c$\\
   (km) & (M$_{\odot}$)  &  ($10^{10}$ G) &  (M$_{\odot}$)\\

\hline                        

 8 &   $1.70 \pm 0.11$ & $10.2 \pm 0.3$ & $0.46 \pm 0.02$\\
  
 8.5   & $1.71 \pm 0.12 $ &$9.7 \pm 0.3$& $0.46 \pm 0.02$ \\
 
 9  & $1.71 \pm 0.12 $ & $9.4 \pm 0.3$& $0.46 \pm 0.02$\\
   
 9.5 & $1.70 \pm 0.13$ & $9.0 \pm 0.3$& $0.46 \pm 0.02$\\   
 10  & $1.69 \pm 0.13$ & $8.8 \pm 0.3$& $0.46 \pm 0.02$\\   

10.5 &  $1.67 \pm 0.14$ & $8.5 \pm 0.3$& $0.45 \pm 0.03$\\
 
11   & $1.64 \pm 0.14$ &  $8.3 \pm 0.3$ & $0.45 \pm 0.03$\\

11.5 &  $1.61 \pm 0.15$ &  $8.1 \pm 0.3$ & $0.44 \pm 0.03$\\

\hline                                             %inserts single line
\end{tabular}
\end{center}

{\small \sc Note} \footnotesize-Uncertainties are
discussed in the text. For clarity we also show 
the values of M$_{NS}$, B, and  M$_c$ for  R$_{NS} =10$ km. The errors
are at 68\% confidence level. 
\end{table}

Assuming different NS radii for Eq.  \ref{Pdot}, we obtain 
different values of the NS mass. We show the NS masses for several
values of the NS radius ranging from 8 to 11.5 km in
Table \ref{TAB_M_R}. The NS radius in X1822-371 cannot be larger than 11.5 km
because the NS mass would be lower  than 1.61 M$_{\odot}$ which is the lower
limit given by \cite{munoz_2005}. This result further constrains the
NS mass range between 1.46 and 1.81 M$_{\odot}$. The NS B-field
ranges between 7.8 and 10.5 $\times 10^{10}$ G, and finally,  
the companion star mass ranges between 0.41 and 0.48 
M$_{\odot}$.

We compared the results in Table \ref{TAB_M_R} with those obtained by
\cite{steiner_10}, which determined an empirical dense matter equation
of state from a heterogeneous data set of six neutron stars: three
Type-I X-ray busters with photospheric radius expansion and three
transient low-mass X-ray binaries. Our comparison was done with the
results reported in Table 7 by \cite{steiner_10}. They are valid for a NS
radius equal to the photospheric radius of the NS.  The authors find
that for a NS mass of 1.6, 1.7, and 1.8 M$_{\odot}$, the corresponding
radius is $10.8^{+0.6}_{-0.9}$, $10.7^{+0.8}_{-1.2}$, and
$10.7^{+0.7}_{-1.1}$ km, respectively, with the errors at 95\%. If the
NS in X1822-371 is similar to those of the sample studied by
\cite{steiner_10}, we can exclude from Table \ref{TAB_M_R} the solutions
for NS radii smaller than 9.5 km. This implies
that the NS mass range is between $1.61 \pm 0.15$ and $1.70 \pm 0.13$
M$_{\odot}$, that the NS B-field in units of $10^{10}$ G is between $8.1
\pm 0.3$ and $9.0 \pm 0.3$, and finally that the companion star mass is
between $0.44 \pm 0.03$ and $0.46 \pm 0.02$ M$_{\odot}$.

 We note that the estimation of the CRSF energy is model
  dependent. The CRSF energy is 0.72 and 0.68 keV, adopting the {\tt
    gabs} and {\tt cyclabs} components, respectively, to fit the averaged
  spectrum.  However, the values of the NS mass, NS magnetic-field  strength,
  and companion star mass at different NS radii are the same as shown in
  Table \ref{TAB_M_R}  even using the CRSF energy obtained from the {\tt
    cyclabs} component, since  this energy and the NS magnetic-field  strength, 
    $B_{12}$, are linearly dependent (see Eq. \ref{B_field})
  and because the spin period derivative weakly depends on
  $B_{12}$ (see Eq. \ref{Pdot}).

The {\tt gabs} component allows us to estimate the temperature 
of the plasma where the CRSF originates, assuming that the broadening
of the line has a thermal origin. 
At the cyclotron resonance
frequency $\omega_c$, electrons at rest absorb photons of energy
$\hbar \omega_c$. For  thermal  Doppler broadening,
$\Delta \omega_D$ is predicted to be \citep{mezaros_92}

$$
\frac{\Delta \omega_D}{\omega_c} = \left(
\frac{2kT}{m_ec^2}\right)^{1/2} |\cos \theta|,
$$
where $\hbar \Delta \omega_D = \sigma_{\rm gabs}$, $\hbar \omega_c =
E_{\rm gabs}$, $ kT$ is the electron temperature, and $m_ec^2$ is the
electron rest energy. The angle $\theta$ measures the direction of the
magnetic field with respect to the line of sight.  Outside the range
$\omega_c \pm \Delta \omega_D$, the cyclotron absorption coefficient
decays exponentially, and other radiative processes become important.
Substituting the values of $\sigma_{{\rm gabs}}$ and $E_{{\rm gabs}}$
shown in Table \ref{TAB_Continuum} (for MOS12 spectrum ranging between
1.5 and 10 keV), we obtain a lower limit on the plasma temperature of
$kT = 8 \pm 2$ keV with the error at 68\% confidence level, which is a factor of
two or three larger than the electron temperature of the {\tt Comptt}
component, but the values are consistent at the $2\sigma$ 
  level. This suggests that the Comptonised component is probably
  produced in the accretion column onto the NS magnetic caps.

We do not observe cyclotron harmonics in the spectrum. To date, the
lowest energy measured for a CRSF produced by electron motion around
the NS magnetic-field lines is 9 keV in the source XMMU
J054134.7-682550 \citep{Mano}; also in that case, harmonics are not
visible in the spectrum. The cyclotron line observed in the spectrum
of the Be/X-Ray Binary Swift J1626.6-5156 has an energy of 10 keV and
only a weak indication of a harmonic at 19 keV
\citep[see][]{decesar}. The source KS 1947+300 has been recently
observed with Nuclear Spectroscopy Telescope Array ({\it NuSTAR}) and
Swift/XRT in the 0.8-79 keV energy range \citep{furst}; a CRSF at 12.5
keV has been observed but no harmonics have been detected.  Finally,
the anomalous X-ray pulsar SGR 0418+5729 shows a CRSF produced by
proton motion, with centroid  at 1 keV, and no harmonics are observed
\citep[see][]{tiengo}.  To now, only the isolated NS CCO 1E1207.4-5209
shows a CRSF also and its first harmonic at 0.7 and 1.4 keV,
respectively \citep[][]{Sanwal_2002,Mereghetti_2002}.  These results
show that, although peculiar, it is possible to observe only the
fundamental harmonic of the CRSF.

Finally, we note that the presence of a CRSF at 0.7 keV shown in this
work contrasts with the recent result suggested by
\cite{sasano_13}, who find a CRSF at 33 keV when analysing the same Suzaku
data as presented in this work.  First of all, we note that a CRSF at 33
keV is detectable including the HXD/PIN data up to 40 keV. However, we
have  shown that the HXD/PIN source spectrum is overwhelmed by
the NXB+CXB spectrum at energies higher than 36 keV (see
Fig. \ref{backpin}).  Furthermore, we note that, at 33 keV,
 the HXD/PIN effective
area  is nearly 50 cm$^2$, while the HPGSPC and PDS instruments on-board BeppoSAX had an effective area of $\sim 200$ and $\sim 500$
cm$^2$, respectively. \cite{Iaria2001_1822} analysed a broad band
spectrum of X1822-371 using the narrow-field instruments on-board
BeppoSAX and did not find any evidence of a CRSF at 30 keV with a
PDS exposure time of 18.7 ks.  The HXD/PIN spectrum has an exposure
time of 37.7 ks, which is a factor of two longer than the PDS exposure
times, but it has an effective area a factor of 10 smaller at 33 keV.
 The presence of a CRSF at 33 keV in the Suzaku data is therefore unrealistic when 
assuming that it does not change in time.

  To verify the presence of a CRSF at 33 keV, we have also analysed 
 the  IBIS and JEM-X  spectra for  an effective
  dead-time-corrected exposure of 874 ks and 283 ks, respectively. 
We fitted both the spectra, adopting the same model as used to fit the XMM-Newton/Suzaku
data and found  no evidence of CRSF  at 33 keV.

We also note that a CRSF at 33 keV is not consistent with the observed
spin-up of the NS.  Assuming that the NS is spinning-up and using the
NS B-field value inferred by a CRSF at 33 keV, \cite{sasano_13}
obtain an intrinsic luminosity of the system of $\sim 3 \times
10^{37}$ erg s$^{-1}$ needed to have the measured spin-up rate.  In
case of spin-up, we expect that the corotation radius $r_c$ has to be
larger than the magnetospheric radius $r_m$.  The values of $r_c$ is
1~180 km and 1~300 km for a NS mass of 1.4 and 2 M$_{\odot}$,
respectively, assuming the value reported by
\cite{sasano_13} as spin period. To estimate the magnetospheric radius, we used the
Eq. \ref{alven} assuming a NS radius of 10 km and a NS B-field of $2.8
\times 10^{12}$ G \citep[see][]{sasano_13}.  For a NS mass of
1.4 M$_{\odot}$, we find that $r_A \simeq 5~940$ km and $r_m = \phi r_A \simeq
3~000$ km, with $\phi=0.5$. We infer that $r_m \simeq 3 r_c$, and this
implies that matter cannot accrete onto the NS; this scenario
corresponds to that discussed by \cite{Ghosh_lamb} for
$\omega_s>>\omega_{max}$ with $\omega_{max}\simeq 0.95$. In fact,
$$
\omega_s \simeq 1.35 B_{12}^{6/7} m^{-2/7}P_s^{-1} L_{37}^{-3/7}
$$
for a NS radius of 10 km \citep[see Eqs. 16 and 18 in][]{Ghosh_lamb}.
Using the values of luminosity, spin period, and NS B-field reported by
\cite{sasano_13}, we find that $\omega_s \simeq 3.2$ for  a NS mass of 1.3 M$_{\odot}$ and $\simeq 2.5$
for 3 M$_{\odot}$. This result
suggests that the scenario is not self-consistent because the values
of luminosity and NS B-field would contradict  the observed spin-up  of the NS.

\section{Conclusion}
We analysed the broadband X-ray spectrum of X1822-371 using all the
presently available data sets that allow for high-resolution
spectroscopic studies. Our aim was to understand 
the nature of the residuals between 0.6 and 0.8 keV
 previously observed in the XMM/EPIC-pn data by
\cite{iaria_2013}. Initially, we analysed the averaged
spectra simultaneously. The broad band spectrum obtained from Suzaku
constrains the continuum emission  well.  The adopted model was a Comptonised
component absorbed by the neutral interstellar matter and partially
absorbed by local neutral matter. We took the Thomson scattering
into account and fixed the value of the equivalent hydrogen column
density of the local neutral matter to that producing the Thomson
scattering.

The values of the parameters are similar for the three data sets and
consistent with previous studies \citep[see e.g.][]{iaria_2013}.  To
fit the residuals between 0.6 and 0.8 keV, we added an absorption
feature with Gaussian profile ({\tt gabs} in XSPEC). Alternatively, we
 adopted an absorbing feature with Lorentzian profile ({\tt
  cyclabs} in XSPEC). In both cases the addition of a CRSF to the
model improved the fit. We found that the improvement does not 
depend sensitively on the exact shape used to model the absorption profile.

 We also detected the spin period of X1822-371 in the EPIC-pn data. We
 obtained the value of  0.5928850(6) s. Using all the measurements
known of the spin period of X1822-371, we  estimated that the
spin period derivative of the source is $-2.55(3) \times 10^{-12}$
s/s, and this confirms  that the neutron star is spinning up. Folding the 
EPIC-pn light curve, we derived a pulse fraction of  0.75\% in the 2-5.4 keV 
energy band.

Using the best-fit values of
the CRSF parameters,  under the assumption that the system is
  accreting at the Eddington limit, we estimate a NS B-field between
$(8.1 \pm 0.3) \times 10^{10} $ and $(9.0  \pm 0.3) \times 10^{10} $ G
for a NS radius ranging between 9.5 and 11.5 km.  
We subsequently constrain the NS mass
assuming that the CRSF is produced at the NS surface. We find that,
for a Gaussian profile of the CRSF, the NS mass is between $1.61 \pm
0.15$ and $1.70 \pm 0.13$ M$_{\odot}$.  The companion star mass is
constrained between $0.44 \pm 0.03$ and  $0.46 \pm 0.02$ M$_{\odot}$.

  Finally, we note that our conclusions contrast with the
  recent results reported by \cite{sasano_13}, who report
  detecting a CRSF at 33 keV (and a corresponding NS-B field of 3
  $\times 10^{12}$ G) in the Suzaku data also used in this work.  To
  address this point, we have selected the whole IBIS and JEM-X public
  data set of the X1822-371 region. We extracted the JEM-X and
  IBIS spectra of X1822-371 having an exposure time of 330 and 874 ks,
  respectively. The INTEGRAL spectrum combined with the XMM, Suzaku,
  and Chandra spectra does not show a CRSF at 33 keV. Furthermore, we
  also show from theoretical arguments that a CRSF at 33 keV is not
  consistent with the evidence that the NS in X1822-371 is
  spinning up.

\begin{acknowledgements} 
The High-Energy Astrophysics Group of Palermo acknowledges support
from the Fondo Finalizzato alla Ricerca (FFR) 2012/13, project N.
2012-ATE-0390, founded by the University of Palermo. This work was
partially supported by the Regione Autonoma della Sardegna through
POR-FSE Sardegna 2007-2013, L.R. 7/2007, Progetti di Ricerca di
Base e Orientata, Project N. CRP-60529, and by the INAF/PRIN 2012-6.
  AR gratefully acknowledges Sardinia Regional Government for the
  financial support (P.O.R. Sardegna F.S.E. Operational Programme of
  the Autonomous Region of Sardinia, European Social Fund 2007-2013 -
  Axis IV Human Resources, Objective l.3, Line of Activity l.3.1.
 \end{acknowledgements} 
\bibliographystyle{aa} % style aa.bst
\bibliography{citations}
\end{document}